\documentclass[pra, notitlepage, showpacs, floatfix, reprint, nobalancelastpage,citeautoscript]{revtex4-1}

\usepackage{graphicx}
\usepackage{multirow}
\usepackage{verbatim, bm}
\usepackage{physics}
\usepackage{textcomp}
\usepackage{cancel}
\usepackage{nicefrac}
\usepackage{mathrsfs}  
\usepackage[normalem]{ulem}
\usepackage{mathtools}
\usepackage{amsfonts}

\usepackage{color}
\usepackage{amsmath}
\usepackage{bbm}
\usepackage{amssymb}
\usepackage{wrapfig}
\usepackage{array, makecell}
\usepackage{pifont}
\newcommand{\cmark}{\ding{51}}
\newcommand{\xmark}{\ding{55}}
\newcommand{\equref}[1]{Eq.~(\ref{#1})}
\newcommand{\pdagger}{{\phantom{\dag}}}
\newcommand{\figref}[1]{Fig.~\ref{#1}}
\usepackage[colorlinks=true]{hyperref}  
\hypersetup{
    bookmarks=true,         
    unicode=false,          
    pdftoolbar=true,        
    pdfmenubar=true,        
    pdffitwindow=false,     
    pdfstartview={FitH},    
    pdfsubject={},   
    pdfcreator={},   
    pdfproducer={}, 
    pdfkeywords={} {} {}, 
    pdfnewwindow=true,      
    colorlinks=true,       
    linkcolor=blue, 
    citecolor=blue,        
    filecolor=magenta,      
    urlcolor=blue           
} 

\begin{document}

\title{Protection of parity-time symmetry in topological many-body systems: \\ non-Hermitian toric code and fracton models}
    \author{Henry Shackleton}
    \affiliation{Department of Physics, Harvard University, Cambridge MA 02138, USA}
        \author{Mathias S. Scheurer}
    \affiliation{Department of Physics, Harvard University, Cambridge MA 02138, USA}
\begin{abstract}
In the study of $\mathcal{P}\mathcal{T}$-symmetric quantum systems with non-Hermitian perturbations, one of the most important questions is whether eigenvalues stay real or whether $\mathcal{P}\mathcal{T}$-symmetry is spontaneously broken when eigenvalues meet. A particularly interesting set of eigenstates is provided by the degenerate ground-state subspace of systems with topological order. In this paper, we present simple criteria that guarantee the protection of $\mathcal{P}\mathcal{T}$-symmetry and, thus, the reality of the eigenvalues in topological many-body systems. We formulate these criteria in both geometric and algebraic form, and demonstrate them using the toric code and several different fracton models as examples. Our analysis reveals that $\mathcal{P}\mathcal{T}$-symmetry is robust against a remarkably large class of non-Hermitian perturbations in these models; this is particularly striking in the case of fracton models due to the exponentially large number of degenerate states.
\end{abstract}

\maketitle
Isolated systems are governed by Hermitian Hamiltonians, with real energy eigenvalues and unitary time evolution. Nonetheless, non-Hermitian Hamiltonians \cite{Bender_2007,Bender_2015,Bender1997,Rotter2009,2020arXiv200601837A}, for which eigenvalues may
generally be complex, are also physically relevant as effective descriptions of a large variety of different systems. For instance, they have been studied in the context of biological~\cite{Nelson1998, Amir2016,Murugan2017}, mechanical~\cite{MechanicalSystem}, and photonic~\cite{Ruter2010, Brandstetter2014, Peng2014, peng2014a, Chang2014, Lin2011, Feng2013,Hodaei2014, Feng2014, Regensburger2012, Peng2016, Gao2015, Xu2016,Jing2015,Zhang2016, Sun2014, Chong2011, Ramezani2010, Guo2009, Klaiman2008, Wu2020,Weidemann311,Naghiloo2019} systems, electrical circuits \cite{EC1,EC2,ExperimentRonny}, cavities \cite{SchaeferMicrowave,SchaeferMicrowave2,Lee2014,Choi2010}, optical lattices \cite{Lee2014a}, superconductors \cite{SCs1,SCs2}, and open quantum systems~\cite{Celardo2009,Rotter2009, Diehl2011, San-Jose2016, Nobuyuki2019, Simon2019, Naghiloo2019}. In the latter case, the emergence of complex eigenvalues
can be interpreted as arising due to dissipation.
On top of a complex spectrum, non-orthogonal eigenstates and exceptional points are unique features of non-Hermitian Hamiltonians, with crucial physical consequences \cite{kato1966perturbation,2012JPhA...45R4016H,2014JPhA...47c5305B}.
In the past few years, there has been growing interest in the condensed matter community in studying non-Hermitian generalizations of quantum
many-body systems. Most of these recent efforts were motivated by the question of how to generalize topological band theory to non-Hermitian systems \cite{ReviewCorrelated,ReviewBands}, uncovering a modified bulk-boundary correspondence \cite{Hu2011,Esaki2011,San-Jose2016,PhysRevLett.116.133903,Yao2018,Kunst2018,Xiong2018,RobertTopolBands,2020arXiv200401886K} and topological classification \cite{Gong2018, Kawabata2018,Lieu2018,Leykam2017,Shen,ZhouLee,Longhi2019,Yuce2019}, as well as exceptional nodal phases \cite{Kozi2017,Zhou1009,Moors2019,Budich2019,Okugawa2019,Rui2019}. 
Furthermore, there has also been research on disordered systems \cite{Disorder2,Disorder3,NonHermitianDisorder,Anderson4} and studies of non-Hermitian physics where many-body correlations play a crucial role, such as non-Hermitian fractional quantum Hall phases \cite{FQHYoshida}, Kondo physics \cite{PhysRevLett.121.203001}, critical points \cite{UedaCritical1,Littlewood}, and many more \cite{PiazzaQMB,SkinSuperfluid,UedaPhaseTransition,Guo2019,Guo2020,FermionicChain,Superfluidity1,Superfluidity2,BoseMottInsulator,TopMottIns1D}.

Among these models, a particularly important and commonly studied class of non-Hermitian Hamiltonians is provided by $\mathcal{P}\mathcal{T}$-symmetric Hamiltonians which are invariant under a combination of parity and time-reversal. Despite being non-Hermitian, these Hamiltonians can exhibit real spectra~\cite{Bender1997, Bender1998, Bender2003,Bender_2007,Bender_2015}. Intuitively, this may be attributed to a balance of gain and loss between the system and its environment. Mathematically, the protection is related to the fact that $\mathcal{P}\mathcal{T}$ symmetry implies that eigenvalues come in complex-conjugate pairs such that isolated real eigenvalues cannot become complex immediately. When they ``meet'' with another eigenvalue, they can either stay on the real axis or form complex-conjugate partners; when the latter happens, $\mathcal{P}\mathcal{T}$ is said to be broken. 
Therefore, the analysis of $\mathcal{P}\mathcal{T}$-symmetry breaking is particularly subtle in systems with (approximate) degeneracies. 

For symmetry-imposed degeneracies, the reality of the eigenvalues can be simply protected by the symmetry itself and the fact that eigenvalues must come in complex conjugate partners. \textit{A priori}, this is different for degeneracies related to intrinsic topological order \cite{WenFQH,SubirReview}: for instance, the toric code model \cite{Kitaev1997} has four ground states on a torus, that are guaranteed to be (exponentially) close in energy, even if all unitary symmetries are broken; similar statements apply to other spin-liquid phases. An even more dramatic ground-state degeneracy (GSD), that scales exponentially with linear system size, is realized in fracton models---novel quantum states of matter that are characterized by excitations with restricted mobility \cite{NandkishoreHermele,PretkoReview}. Similar to spin-liquids, the GSD of fracton phases is topological in the sense that the different ground states are locally indistinguishable. One might be tempted to conclude that turning on a non-Hermitian, $\mathcal{P}\mathcal{T}$-symmetric perturbation in such systems will immediately lead to complex ground-state energies. Contrary to these expectations, we demonstrate in this paper that the reality of the ground-state eigenvalues in these phases can be surprisingly robust against a large class of such perturbations, even if all unitary symmetries are broken and in the presence of exponentially many degenerate states. 

More specifically, we study under which conditions the eigenvalues of a given (almost degenerate) subspace of a Hermitian quantum system will stay real upon adiabatically turning on a non-Hermitian perturbation such that the total Hamiltonian commutes with a generalized $\mathcal{P}\mathcal{T}$ symmetry. Here, ``adiabatically'' refers to keeping the gap to all other states finite and ``generalized $\mathcal{P}\mathcal{T}$'' indicates that $\mathcal{P}$ does not have to be spatial inversion, but might be any unitary operator. We first discuss a general mathematical condition for the eigenvalues to stay real and, hence, $\mathcal{P}\mathcal{T}$ symmetry to be protected.  
We then demonstrate that this condition has strong implications for the protection of $\mathcal{P}\mathcal{T}$ symmetry in the ground-state manifold of systems with topological GSDs, taking the toric code \cite{Kitaev1997}, the X-cube model \cite{Vijay2016}, the checkerboard models \cite{Vijay2016,VijayHaahFu1}, Haah's 17 CSS cubic codes~\cite{Haah2011}, and the large class of quantum fractal liquids of Ref.~\cite{Yoshida2013} as examples. It is found that $\mathcal{P}\mathcal{T}$ symmetry will be preserved on systems with even linear system sizes, $L_j$, (in some Haah codes, divisibility by $4$ is required) for a large class of perturbations, while it is generically fragile in systems with odd $L_j$. 

We emphasize that understanding the preservation or breaking of $\mathcal{P}\mathcal{T}$ symmetry is not only one of the central theoretical questions of $\mathcal{P}\mathcal{T}$-symmetric quantum mechanics, but also of practical relevance for experimental realizations and potential applications of effectively non-Hermitian systems. We hope that our framework for predicting the stability of the reality of eigenvalues and the presence or absence of exceptional points will provide greater control over the effects of non-Hermitian perturbations, which is, e.g., important for the observation of power-law oscillations \cite{Jorg2019, Takasu2020,Regensburger2012,LeiPTQuantumDynamics} and the potential applications as topological lasers \cite{TopologicalLaser1,TopologicalLaser2,TopologicalLaser3} and sensing devices \cite{WiersigSensing,LiuSensing}.

The remainder of the paper is organized as follows. In Sec.~\ref{GeneralFormOfPerturbations}, we define the type of non-Hermitian Hamiltonians we are interested in,  $\mathcal{P}\mathcal{T}$ symmetry, and the more general condition of pseudo-Hermiticity. We also discuss the general, mathematical condition for colliding eigenvalues to stay real. It is first applied to the toric code, in Sec.~\ref{ToricCode}, to the X-cube, checkerboard models, and Haah's codes in Sec.~\ref{FractonModels}, and finally to the fractal liquids of Ref.~\cite{Yoshida2013} in Sec.~\ref{FractalLiquids}. A summary of our findings is provided in Sec.~\ref{Conclusion}.

\section{Pseudo-Hermitian Perturbations}\label{GeneralFormOfPerturbations}
We start with a general explanation of the class of non-Hermitian perturbations under consideration.
To this end, let us first assume that our non-Hermitian Hamiltonian $H$ admits a complete biorthonormal eigenbasis $\{\ket{\psi}, \ket{\phi}\}$ \cite{2014JPhA...47c5305B}, which means that 
  \begin{align}\begin{split}
    H \ket{\psi_n} &= E_n \ket{\psi_n}\,,
    \\
    H^\dagger \ket{\phi_n} &= E_n^* \ket{\phi_n}\,,
    \\
    \braket{\phi_n}{\psi_m} &= \delta_{mn}\,.
  \label{BiorthogonalBasis}\end{split}\end{align}
This is equivalent to the statement that $H$ is diagonalizable, which is a very natural assumption for a generic (non-Hermitian) Hamiltonian of a physical system. Note, however, that it can be violated, most importantly at exceptional points \cite{kato1966perturbation,2012JPhA...45R4016H}, which we will discuss separately below. 

In the study of non-Hermitian perturbations to quantum systems, 
it is common to further assume that these Hamiltonians are $\mathcal{P}\mathcal{T}$-symmetric~\cite{Bender1997, Bender1998, Bender2003,Bender_2007,Bender_2015}, i.e., $[H,\mathcal{P}\mathcal{T}] = 0$, where $\mathcal{P}$
can be abstractly defined as any unitary operator that squares to $\mathbbm{1}$, and $\mathcal{T}$
is complex conjugation in a certain basis. Doing so imposes additional restrictions on the
spectrum of $H$. Eigenvalues must come in complex conjugate pairs, as $H (\mathcal{P} \mathcal{T}) \ket{\psi_n}
= E_n^* (\mathcal{P} \mathcal{T}) \ket{\psi_n}$. Importantly, this means that if one starts with a Hermitian,
$\mathcal{P} \mathcal{T}$-symmetric Hamiltonian and applies a $\mathcal{P}\mathcal{T}$-symmetric
non-Hermitian perturbation, isolated eigenvalues cannot become complex on their own---they must merge
with another eigenvalue on the real axis before becoming complex. This feature leads to the reality of
energy spectra generally being robust to sufficiently small $\mathcal{P}\mathcal{T}$-symmetric perturbations,
although degenerate subspaces are not necessarily protected from becoming complex.
When $\mathcal{P} \mathcal{T} \ket{\psi_n} \propto \ket{\psi_n}$, $\mathcal{P} \mathcal{T}$ symmetry is said to be ``unbroken'' and the associated eigenvalues are real. Once eigenvalues meet and become complex, $\mathcal{P} \mathcal{T}$ symmetry is ``broken'' and $\ket{\psi_n}$ is not an eigenstate of $\mathcal{P} \mathcal{T}$ any more.

In this work, however, we do not restrict ourselves to $\mathcal{P}\mathcal{T}$ symmetry, 
and instead impose a closely related but more general condition of \textit{pseudo-Hermiticity}~\cite{Mostafazadeh2001, 
Mostafazadeh2001a, Mostafazadeh2002}.
A Hamiltonian $H$ is pseudo-Hermitian if there exists a linear operator $\eta$, which we will refer to as the \textit{metric operator}, such that
\begin{subequations}\begin{equation}
    \eta H \eta^{-1} = H^\dagger. \label{PseudoHermiticity}
\end{equation}
In this paper, we take $H$ to consist of a Hermitian component, $H_0$,
and a non-Hermitian perturbation, $\epsilon V$, with magnitude that we control with $\epsilon \in \mathbbm{R}$: 
\begin{equation}
H=H_0 + \epsilon \, V, \quad H_0^\dagger = H_0. \label{GeneralFormOfHam}
\end{equation}\label{CompleteDesciptionOfHam}\end{subequations}
For the Hermitian part $H_0$, \equref{PseudoHermiticity} implies $\comm{\eta}{H_0} = 0$, i.e., 
$\eta$ is a symmetry of the unperturbed Hamiltonian. We also take $\eta$ to be unitary, so that \equref{PseudoHermiticity} is equivalent to
$\eta H \eta^\dagger = H^\dagger$. The purpose of this work is to derive and discuss general conditions under which (certain physically relevant parts of) the spectrum of $H$ in \equref{GeneralFormOfHam} can stay real upon adiabatically turning on $\epsilon$.

This condition of pseudo-Hermiticity (\ref{PseudoHermiticity})
is manifestly identical to $\mathcal{P}\mathcal{T}$ symmetry with $\mathcal{P} \equiv \eta^{-1}$
provided $H$ is symmetric, $H = H^T$. In fact, it was shown \cite{2018arXiv180101676Z} that any $\mathcal{P}\mathcal{T}$ symmetric, finite-dimensional Hamiltonian is also pseudo-Hermitian.
For this reason and since it does not involve any anti-linear operators and, thus, does not require a choice of basis, we focus on pseudo-Hermiticity in this work. Moreover, pseudo-Hermiticity
gives a more systematic way of constructing non-Hermitian
perturbations $\epsilon\,V$ to Hermitian models: one can immediately 
obtain all the possible choices of $\eta$ as it has to be a symmetry of the unperturbed, Hermitian part, $H_0$, of the model, which then
specifies the suitable non-Hermitian perturbations.

\subsection{Protection of reality of energies}
If $H$ is pseudo-Hermitian, complex eigenvalues also must come in conjugate pairs, since the combination of Eqs.~(\ref{BiorthogonalBasis}) and (\ref{PseudoHermiticity}) implies
$H \eta^{-1} \ket{\phi_n} = E_n^* \eta^{-1} \ket{\phi_n}$. As is the case
with $\mathcal{P}\mathcal{T}$-symmetric perturbations, this means that
if a non-Hermitian perturbation is applied, the reality of isolated eigenvalues 
is stable to small pseudo-Hermitian perturbations. If a group
of eigenvalues are degenerate (or almost degenerate) under $H_0$---as is common in models involving symmetries or topological superselection sectors---they are generally not stable to pseudo-Hermitian perturbations.
In these cases, we identify two main mechanisms by which these degenerate 
eigenvalues can stay real under pseudo-Hermitian perturbations: 

\textit{(I)} The first method of ensuring degenerate eigenvalues stay real
is simply to preserve the degeneracy under pseudo-Hermitian 
perturbations. Pseudo-Hermiticity implies that if degenerate
eigenstates are going to become complex, they must acquire
imaginary parts with opposite signs. If one forces the (in general complex)
eigenvalues to remain degenerate, this can never be satisfied
for a non-zero imaginary component, unless the eigenvalues meet with another set of symmetry-unrelated eigenvalues. The latter, however, requires a sufficiently large value of $\epsilon$, as symmetry-unrelated states are generically not degenerate for $\epsilon=0$. The symmetries enforcing the degeneracy can be unitary symmetries, fermionic time-reversal symmetry \cite{2020arXiv200401886K}, or even bosonic time-reversal symmetries unique
to pseudo-Hermitian systems~\cite{Sato2011}. 

\textit{(II)} The second mechanism is more subtle and our main focus in this work. 
If a pseudo-Hermitian term breaks all symmetries protecting the degeneracy,
the eigenvalue splitting will generally be nonzero. This splitting can be
either real or imaginary. However, one can show that \textit{if all
the eigenstates of $H_0$ of the degenerate (or almost degenerate) subspace of interest have the same eigenvalue under $\eta$,
then this splitting will always be real.}
This mathematical fact can be readily understood within the framework of $G$-Hamiltonian systems developed by Krein, Gel'fand and Lidskii \cite{krein1950generalization,gel1955structure} in the 1950s for the case of Hermitian $\eta$. In Appendix \ref{ap:perturbation}, we provide a simple and physically insightful proof to all orders of perturbation theory that works for $\eta$ being Hermitian or unitary.
Furthermore, our analysis shows that\textit{, if the eigenvalues of $\eta$ are identical, the projections of the associated eigenstates to the (almost) degenerate subspace of $H_0$ will be orthogonal to first order in $\epsilon$ and to zeroth order in the limit of a large gap to the rest of the spectrum}; it also follows that, as long as the energetic separation of the subspace of interest from the rest of the spectrum is sufficiently large, they will be approximately orthogonal in the entire Hilbert space, even though the Hamiltonian is not Hermitian any more. 
This is very different when the eigenvalues of $\eta$ are not the same. In that case, there can be exceptional points \cite{kato1966perturbation,2012JPhA...45R4016H}, where the Hamiltonian is defective, eigenstates coalesce and become identical, irrespective of how large the gap to the other states of the system is.

Intuitively, this is related to the fact that the Hamiltonian restricted to the degenerate subspace is Hermitian: denoting the degenerate eigenfunctions by $\ket{\psi_i}$ and writing $\eta \ket{\psi_i} = e^{i \delta} \ket{\psi_i}$,  we have
\begin{equation}
    \hat{H}_{ij} \equiv \bra{\psi_i} H \ket{\psi_j} \stackrel{(\ref{PseudoHermiticity})}{=} \bra{\psi_i} \eta^{-1} H^\dagger \eta \ket{\psi_j} = \hat{H}^*_{ji}. \label{HermitianMatrixInEigenspace}
\end{equation}
The complete argument of Appendix \ref{ap:perturbation} involves constructing a full effective Hamiltonian for the degenerate subspace, and showing that it is Hermitian via similar reasoning.

\subsection{Remarks on condition for reality}
Some comments should be made regarding the possibility of multiple degeneracies
and multiple metric operators.
In the case of a two-fold degeneracy, i.e., two eigenvalues being identical, there are only two possibilities for the eigenvalues of $\eta$---either
they both have the same eigenvalue under $\eta$, or they are different. In the former
case, the splitting is always real. In the latter case, the energy splitting
can be real or complex depending on the magnitudes of the matrix elements in the effective
Hamiltonian.
When there are more than two degenerate eigenstates, the full criteria becomes more complicated,
as some eigenvalues may become complex while others stay real. For a concrete system,
it should always be possible to determine the nature of the splitting through perturbation
theory, using the methods described in Appendix~\ref{ap:perturbation}. 
However, we note that it is \textit{always}
the case that if all the unperturbed eigenstates have the same eigenvalue under $\eta$,
their energies will stay real.

One can also consider a case where there is a two-fold degeneracy, but multiple possible
choices of metric operators. If there are two metric operators, $\eta_1$ and $\eta_2$,
such that both eigenstates have the same eigenvalue under $\eta_1$
and different eigenvalues under $\eta_2$, the degeneracy can be protected as a consequence of the mechanism \textit{(I)} above: $S=\eta^{-1}_1 \eta^{\phantom{-1}}_2$ is, by construction, a symmetry of $H$ and if $S \ket{\psi_1} = \ket{\psi_2}$, the eigenvalues will remain identical for $\epsilon \neq 0$. However, the pseudo-Hermiticity
of $H$ with respect to $\eta_2$ can be broken without causing the eigenvalues to become
complex. 

In this paper, we focus on the protection mechanism \textit{(II)} for the reality of the eigenvalues, i.e., on cases where pseudo-Hermitian perturbations break all the relevant symmetries, eigenvalue degeneracies are not preserved and hence the interplay
between the metric operator and the unperturbed eigenstates are important
in deducing whether the energies stay real.
The general procedure for utilizing this phenomenon goes as follows.
First, specify a subspace of interest, whose energies are separated from the rest of the spectrum. Next, identify
the unitary symmetries under which the subspace has a definite eigenvalue under. These symmetries will yield a class of non-Hermitian perturbations---namely, those that are pseudo-Hermitian with the symmetry as a metric operator---for which the degenerate eigenvalues will stay real.

This notion of stability is useful in quantum systems when the subspace under consideration is well-separated
from the rest of the spectrum. In the remainder of the paper we will be concerned with gapped many-body systems with several degenerate ground states and discuss under which conditions the ground-state energies can remain real, provided the perturbations
do not close the gap between the ground and excited states.

\section{Non-Hermitian Toric Codes}\label{ToricCode}
We begin with a study of non-Hermitian perturbations to the two-dimensional toric code~\cite{Kitaev1997}, 
focusing on the reality of the ground-state subspace. Non-Hermitian
generalizations of the toric code \cite{UedaPhaseTransition,Guo2020} or closely related models \cite{Guo2019} have recently been studied; these works, however, have a different focus and a systematic understanding of the stability of $\mathcal{P}\mathcal{T}$ symmetry or, more generally, of the reality of the spectrum in the ground-state subspace remains unexplored.

\begin{figure}
\begin{center}
 \includegraphics[width=0.3\textwidth]{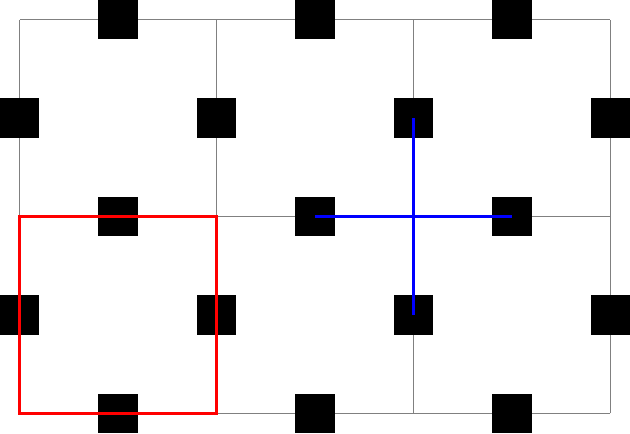}
  \includegraphics[width=0.2\textwidth]{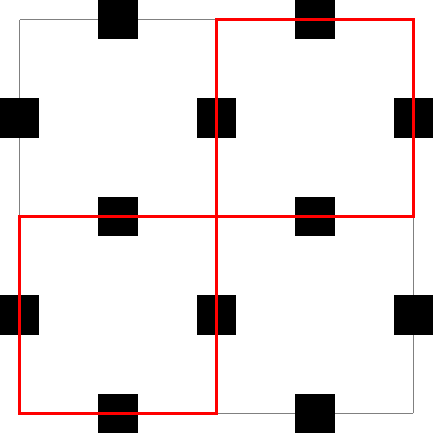}
  \end{center}
  \caption{(Top) The toric code is defined on
  a cubic lattice, with Pauli spins on each edge.
  The Hamiltonian is a sum of stabilizers, 
  consisting of either a product of $X$ operators
  on a plaquette (red), or a product of $Z$ operators
  adjacent to a vertex (blue). We here show the case $L_x=3$ and $L_y=2$. (Bottom) On an even-by-even lattice (depicted here for a $2 \cross 2$ lattice), every site can be covered by
  a combination of non-overlapping plaquette operators (red). The four sites that are seemingly not covered
by plaquette operators are redundant due to periodic boundary conditions. This covering is also possible with vertex operators. The coverings of larger lattices
can be accomplished by sewing together copies of this $2 \cross 2$ covering---of course, this only works
for even-by-even lattices.
\label{fig:tcCovering}}
\end{figure}

The toric code is defined on a square lattice, with Pauli spins
on every edge, see \figref{fig:tcCovering}. We denote the number of sites along the $x$ and $y$ directions by $L_x$ and $L_y$ and focus on periodic boundary conditions. The Hamiltonian is
\begin{equation}
  H^{TC} =  - \alpha \sum_{c} A_c - \beta \sum_{p} B_p, 
  \label{eq:toricCode}
\end{equation}
where we have introduced the vertex operators, $A_c$, which cover the four spins adjacent to a vertex $c$, and the plaquette operators, $B_p$, which cover the four spins on a plaquette $p$,
\begin{equation*}
  \begin{aligned}
    A_c = \prod_{i \in c} Z_i, \quad B_p = \prod_{i \in p} X_i.
  \end{aligned}
\end{equation*}
Unless stated otherwise, we will use $\alpha=\beta=1$. In accordance with quantum code terminology, we refer to
$A_c$ and $B_p$ collectively as ``stabilizers.''

Each term in \equref{eq:toricCode}
commutes with the rest of the Hamiltonian, so the ground states can be obtained by minimizing the energy of each
operator independently. Any state $\ket{\psi}$ in the ground-state subspace satisfies 
$A_c \ket{\psi} = B_p \ket{\psi} = \ket{\psi}$. If defined on a torus, one can define loops of $Z$ or $X$ operators
that wind around either of the two cycles of the torus. These logical string operators, that cannot be
deformed to the identity by applications of stabilizers, imply a fourfold degenerate ground state, with string
operators acting irreducibly within that subspace.

\subsection{Pseudo-Hermitian perturbations}
We are interested in pseudo-Hermitian perturbations to \equref{eq:toricCode} and how they affect the degenerate ground states. 
To this end, let us first focus on three possible choices of $\eta$,
\begin{equation}
    \eta = \prod_i X_i\,, \prod_i Y_i\,, \prod_i Z_i,
\label{OneNaturalChoiceOfEta}\end{equation}
where the product involves all sites of the system, and postpone the discussion of other options to Sec.~\ref{OtherMetricOperators} below. One can easily check that $\comm{H^{TC}}{\eta} = 0$. 

In contrast with many other features of the toric code, which only depend on the topology of the manifold, the eigenvalues of the ground states under $\eta$ in \equref{OneNaturalChoiceOfEta} are highly sensitive to the system size. 
On an even-by-even lattice, the entire ground-state subspace has the 
same eigenvalue under $\eta$. This can most easily be seen by the fact that $\eta$ can be written as a product
of plaquette and vertex operators, and must give eigenvalue $+1$ in the ground-state subspace as a result, 
see Fig.~\ref{fig:tcCovering}. This cannot be accomplished on any other lattice size for the following reason:
a straight line drawn along the $x$ ($y$) direction going through the centers of the plaquettes will intersect $L_x$ ($L_y$) sites. If we attempt to cover
the full lattice with plaquette operators, the placement of an additional operator will always change the number of 
covered sites on the line by an even amount. The same holds true for vertex operators and lines drawn through the vertices. Therefore, if either $L_x$
or $L_y$ is odd, the full lattice can never be assembled solely from stabilizers. The fact that $\eta$ cannot be written as a product of stabilizers is sufficient to show that not all ground states can have the same eigenvalue under $\eta$. To see this, suppose that all ground states have the same eigenvalue under $\eta$. If this holds, then we can add $\eta$ to the group of stabilizers of the toric code without modifying the GSD. If $\eta$ is independent from the rest of the stabilizers, we arrive at a contradiction, since increasing the number of independent stabilizers lowers the GSD.

The observation that all the ground states have the same sign under $\eta$ can also be seen by noting that 
$\eta$ commutes with all the logical string operators on an even-by-even lattice, 
which take the system between different ground states.
On an odd-by-even or an odd-by-odd lattice, $\eta$ anti-commutes with at least one of the logical string
operators, which in both cases lead to two ground states having eigenvalue $+1$ and the other two having eigenvalue $-1$. 

What sort of perturbations, $\epsilon V$, can we add to our Hamiltonian for which $\eta V \eta^\dagger = V^\dagger$?
Writing $V=i \mathcal{O}$, this requires
\begin{equation}
  \eta \mathcal{O} = - \mathcal{O}^\dagger \eta\,,
  \label{eq:ptSymmetryCondition}
\end{equation}
which reduces to $\acomm{\eta}{\mathcal{O}} = 0$ for Hermitian $\mathcal{O}$.
Taking $\eta$ to be the product of $Y$ operators for concreteness, 
this means that $\mathcal{O}$ can be a sum, $\mathcal{O} = \sum_t g_t \mathcal{O}_t$, $g_t\in \mathbbm{R}$, over terms $\mathcal{O}_t$ which are products of Pauli matrices, only constrained to contain an odd number of $X_i$ and $Z_i$. This includes a large class of perturbations such as random, planar fields, $V= i \sum_i (g_{i1} X_i + g_{i3} Z_i)$, $g_{i1},g_{i3}\in \mathbbm{R}$, and highly non-local terms, such as $i \sum_{i<j<k} g_{ijk} X_i X_j X_k$, or $\sum_{i<j<k} g_{ijk} X_i Y_j Y_k$, $g_{ijk}\in\mathbbm{R}$. Since each term satisfies \equref{eq:ptSymmetryCondition} separately, there is no relation between the prefactors of the different terms required and we can think of them as random, non-Hermitian disorder, that in general breaks all symmetries of the system (other than $\mathcal{P}\mathcal{T}$). 

\begin{figure}
  \centering
  \includegraphics[width=\linewidth]{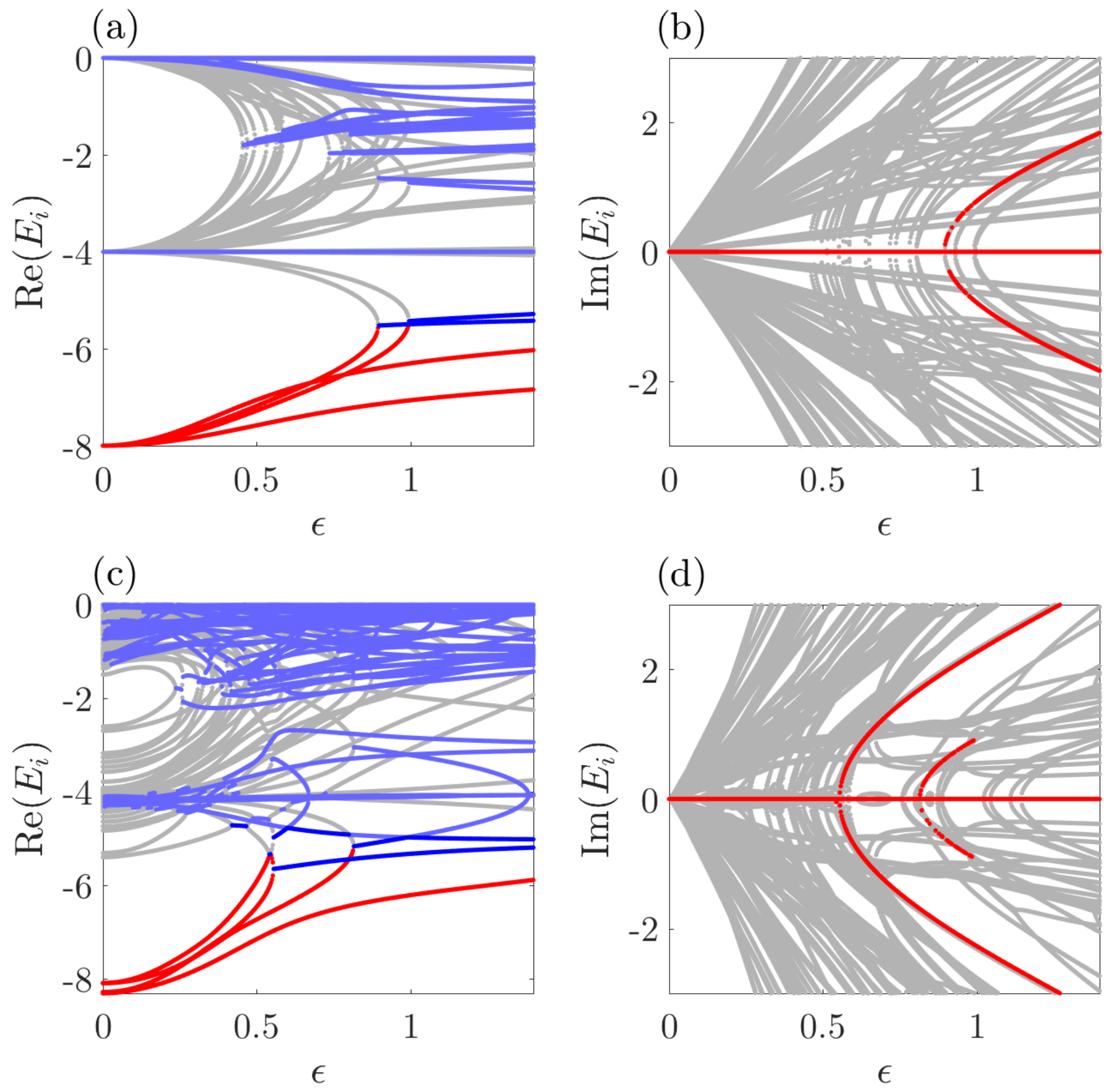}
  \caption{Spectrum of the toric with non-Hermitian random field perturbation, $i \epsilon \sum_i g_i X_i$, where $g_i$ was initialized randomly according to a Gaussian distribution with mean and variance $1$. In (a,b) and (c,d) we take the bare toric-code Hamiltonian (\ref{eq:toricCode}) and the perturbed one, Eq.~(\ref{eq:tcPerturbation}) with Gaussian distributed $h_i$ (mean $0$ and standard deviation $0.4$), as starting point, respectively. In (a,c), the real part of the energy is shown with red and gray referring to real-valued ground and excited energy levels, whereas eigenvalues with a complex part (broken $\mathcal{P}\mathcal{T}$ symmetry) are indicated in blue. The corresponding imaginary parts can be found in (b,d) with red indicating the ground states, defined as those four states with the lowest $\text{Re}(E_i)$.  \label{fig:tcED}}
\end{figure}

In combination with our results of Sec.~\ref{GeneralFormOfPerturbations}, this implies that on an even-by-even lattice, the ground-state subspace of the toric code 
remains real under the large class of pseudo-Hermitian perturbations that satisfy \equref{eq:ptSymmetryCondition} with $\eta$ given by \equref{OneNaturalChoiceOfEta}. As the eigenvalues must stay real for small perturbations, they never exhibit any square root
singularities \cite{2012JPhA...45R4016H} and exceptional points are avoided in the ground-state subspace.
This is verified by exact diagonalization (ED) of the toric code spectrum in Fig.~\ref{fig:tcED}(a,b), where it can be seen that the ground-state energies can only become complex when meeting with the excited states. As such, the $\mathcal{P}\mathcal{T}$ symmetry of the ground-state manifold is protected by the gap to the excited states.

In contrast, on a lattice that is not even-by-even, the ground
states generically become complex immediately upon applying the same non-Hermitian perturbations. This 
sensitivity of the ground state to the system size can be
thought of as representative of the highly entangled nature
of the toric code ground states. Even if one was to consider
arbitrarily large system sizes, the toric code ground states
are still able to ``detect" whether the system size is even or odd.
A similar interpretation of this phenomenon is that even for local
perturbations, the order in perturbation theory in which the ground
state energy splitting will occur necessarily involves a non-local
operator which winds around the torus and, as such, can be
sensitive to (the parity of) the system size. 

This sensitivity to system size may seem surprising, as the toric
code is a paradigmatic example of topological order where physical
features are only sensitive to the genus of the underlying
manifold. To reconcile this, we emphasize that these non-Hermitian features
assume the preservation of $\mathcal{P}\mathcal{T}$-symmetry or
pseudo-Hermiticity, which is a common assumption for non-Hermitian systems. 
Since the topological order of the Hermitian toric code would 
remain even if this symmetry was broken, the phenomenon we observe is not
topological in a strict sense, even though it can be interpreted as arising
due to the long-range entanglement intrinsic to topological order. In 
the following sections, we show that despite this, these features are 
a robust property of the topological phase rather than a fine-tuned 
consequence of the exact solvability of Eq.~(\ref{eq:toricCode}).

\subsection{Starting with perturbed toric code}
One might wonder whether the remarkable protection of $\mathcal{P}\mathcal{T}$ and reality of the ground-state energies is just a consequence of the highly fine-tuned and exactly solvable toric code Hamiltonian (\ref{eq:toricCode}) or a more general property of the underlying topologically ordered phase. 
To investigate this, let us take as our base Hamiltonian the toric code with some small Hermitian perturbation, for example a field along the $Z$-direction with in general spatially varying amplitude,
\begin{equation}
    H_0 = H^{TC} +  \sum_i h_i Z_i\,, \quad h_i \in \mathbbm{R}.
    \label{eq:tcPerturbation}
\end{equation}
The perturbation in \equref{eq:tcPerturbation} forces us to choose $\eta = \prod_i Z_i$ in \equref{OneNaturalChoiceOfEta}, as it is the only one that commutes with the Hermitian Hamiltonian. Note that, of course, a completely random Hermitian field will break all symmetries and no $\eta$ is possible; we are, however, not interested in this case as the Hamiltonian would break $\mathcal{P}\mathcal{T}$ \textit{explicitly} and the question of whether it is broken \textit{spontaneously} would become ill defined.

With the additional perturbation in \equref{eq:tcPerturbation}, we no longer have exactly degenerate ground states for $\epsilon=0$, but a finite energy splitting that is exponentially suppressed by the system size. The four low-energy states will still all be even under $\eta$ for an even-by-even lattice, since our perturbation respects
the $\eta$ symmetry. Consequently, the ground-state energies will stay real, even when they ``meet'' each other at finite $\epsilon$, as long as the gap to excited states stays finite. The protection of $\mathcal{P}\mathcal{T}$ symmetry is, thus, a more general property of the underlying phase with topological order. We also demonstrate this with a concrete example in \figref{fig:tcED}(c,d).

\subsection{Exceptional points}
A surprising observation is that, while these Hermitian perturbations do not
change whether the ground states become complex, they do change the nature of
\textit{how} they become complex. Non-Hermitian Hamiltonians can exhibit
exceptional points \cite{kato1966perturbation,2012JPhA...45R4016H}. Here, the eigenvalues coalesce, the matrix becomes defective, i.e., also the eigenvectors become degenerate, and the eigenvalues exhibit a square-root singularity in the tuning parameter, in our case $\epsilon$, in the sense that the difference of eigenvalues scales with $\sqrt{\epsilon_0-\epsilon}$. This has crucial consequences, e.g., for the Green's function that exhibits a pole of second order in addition to the conventional first-order pole \cite{2012JPhA...45R4016H}.
For pseudo-Hermitian or
$\mathcal{P} \mathcal{T}$-symmetric Hamiltonians, exceptional points 
typically arise at the moment when two eigenvalues meet on the real line 
and become complex. If we start with an unperturbed toric code on an 
even-by-odd lattice and apply a non-Hermitian, pseudo-Hermitian   perturbation $\epsilon V$,
such as an imaginary transverse field, the degenerate ground states can
immediately become complex. However, this degeneracy is not an exceptional
point, since the degeneracy occurs in the Hermitian limit and must admit
a complete basis of eigenvectors. In contrast, if one first applies a 
Hermitian perturbation, such as in \equref{eq:tcPerturbation}, and then
$\epsilon V$, we have verified by ED on a $2 \times 3$ lattice that the ground states
\textit{will} form an exceptional point when they meet each other on the real 
line to become complex, and the corresponding eigenstates become identical. We emphasize that this is true for arbitrarily small Hermitian 
perturbations. This is in stark contrast to systems with even $L_x$, $L_y$; here the ground-state energies must stay real for small perturbations, such that their splitting does not exhibit the characteristic square-root behavior $\propto \sqrt{\epsilon_0-\epsilon}$, and exceptional points do not occur.

To illustrate this subtle behavior of perturbed systems with odd system sizes, let us take a two-level system as an effective description of two ground states with opposite eigenvalue of $\eta$ meeting to become complex. Denoting Pauli matrices acting in this subspace by $\sigma_{x,y,z}$, we have $\eta=\sigma_z$ and the most general pseudo-Hermitian Hamiltonian has the form 
\begin{equation}
    h = E_0 \mathbbm{1} + \Delta \sigma_z + i \epsilon \left( \cos \alpha \, \sigma_x + \sin \alpha \, \sigma_y \right), \label{ModelHamiltonian}
\end{equation}
with the real-valued parameters $E_0$, $\Delta$, $\alpha$, and $\epsilon$; the latter parameterizes the strength of anti-Hermitian perturbations as before. Note that the model is $\mathcal{P}\mathcal{T}$ symmetric only if $2\alpha/\pi \in  \mathbbm{Z}$. The right eigenvalues and eigenvectors of $h$ in \equref{ModelHamiltonian} are given by $E_{\pm} = E_0 \pm \sqrt{\Delta^2 - \epsilon^2}$ and $\psi_{\pm} \propto (\Delta \pm \sqrt{\Delta^2 - \epsilon^2}, i \epsilon \, e^{i\alpha})^T$. The eigenvalues meet when $\epsilon=\pm \Delta$ and become complex for $|\epsilon| > |\Delta|$. When $\Delta = 0$, however, this is not an exceptional point as $\psi_{\pm} \rightarrow (\pm 1,\text{sign}(\epsilon)  e^{i\alpha})^T/\sqrt{2}$, forming an orthonormal basis, and $\Delta E = E_+ - E_- \rightarrow 2 |\epsilon|$ scaling linearly with $\epsilon$, for $\Delta \rightarrow 0$. For $\Delta \neq 0$, instead, we get $\psi_+ \rightarrow \psi_-$ when $\epsilon \rightarrow \pm \Delta$, showing that the matrix becomes defective, and the difference of eigenvalues scales as $\Delta E \sim 2\sqrt{2\epsilon_0}\sqrt{\epsilon_0 - \epsilon}$, for $\epsilon$ near $\epsilon_0 = \pm \Delta$. It is also readily verified that the overlap, $\braket{\phi_{\pm}}{\psi_{\pm}}$, with the corresponding left eigenvector is non-zero except for the exceptional points $\epsilon = \pm \Delta \neq 0$, where it vanishes; this ``self-orthogonality'' rules out the construction of a bi-orthogonal basis as in \equref{BiorthogonalBasis}. In summary, we should think of the special case of vanishing splitting, $\Delta=0$ or of the unperturbed toric code, as a fine-tuned limit where two lines of exceptional points, $\epsilon = \pm \Delta$, meet and give rise to a non-defective Hamiltonian, as required by Hermiticity.

We finally point out that this behavior is also visible on an even-by-even lattice when taking into account the excited states: as can be seen in \figref{fig:tcED}(b,d), the imaginary part of the excited states that become complex at infinitesimal $\epsilon$ scales linearly in $\epsilon$, whereas the $\mathcal{P}\mathcal{T}$ symmetry breaking at finite $\epsilon$ exhibits the aforementioned square-root singularity.

\subsection{Other metric operators}\label{OtherMetricOperators}
So far, we have focused on the three different choices of $\eta$ in \equref{OneNaturalChoiceOfEta}, but there are in principle many more possibilities for the bare toric code model (\ref{eq:toricCode}), as it possesses many other symmetries. Here, we argue that our choices of $\eta$ are unique provided we assume our anti-Hermitian perturbations can be disordered and are not required to have a specific spatial structure.

As a starting point, one might use spatial symmetries---lattice translations $T_{x,y}$, four-fold rotation $C_4$, and inversion $I$ and combinations thereof. For instance, $\eta = I$ with 
\begin{equation}
    I \mathcal{O}_i I^{-1} = \mathcal{O}_{-i}, \quad \mathcal{O}_i = X_i,\,Y_i,\,Z_i, \label{SpatialInversion}
\end{equation}
is clearly a symmetry, $[H^{TC},I]=0$, and it is easy to see that all ground states have the same eigenvalue under it for any system size (the same holds for $T_{x,y}$ but not for $C_4$). However, it is not a natural choice for a generic system with spatially varying Hermitian or non-Hermitian perturbations, such as those discussed above. For example, for an imaginary field, $V=i \sum_i \sum_{\mu=1}^3 g_{i\mu} (X_i,Y_i,Z_i)_\mu$, it would require $g_{i\mu} = - g_{-i\mu}$ and, hence, fine-tuning between spatially distant sites. Not even a site-independent complex field is possible.

Having established that choosing an $\eta$ which relates spatially distant sites requires fine tuning, we focus on $\eta$ that commute with all stabilizers separately. This requirement can alternatively be thought of as a restriction to symmetries that are preserved in the presence of spatial disorder in the couplings of the bare toric code, i.e., $\alpha \rightarrow \alpha_c>0$, $\beta\rightarrow \beta_p>0$ in \equref{eq:toricCode}. This leads to two distinct classes of possible $\eta$, schematically given by 
\begin{subequations}
\begin{equation}
\eta =\prod (\text{stabilizers}) \label{FirstClassOfEta}
\end{equation}
or
\begin{equation}
 \eta =\prod (\text{stabilizers})(\text{logical strings}) , \label{SecondClassOfEta}
\end{equation}\label{ClassesOfEtas}\end{subequations}
where ``logical strings'' stands for strings of $X_i$ or $Z_i$ operators along a non-contractible loop of the torus connecting the different ground states \footnote{Strictly speaking, $\eta$ can also be a sum of terms of the form in \equref{ClassesOfEtas}. While this is already excluded by our assumption of unitary $\eta$, extending to non-unitary $\eta$ would not alter the discussion here. To see this, consider $\eta= \eta_1 + \eta_2$, with $\eta_j$ of the form (\ref{ClassesOfEtas}), and anti-Hermitian $V$. In order for this to admit a pseudo-Hermitian perturbation that would not have been allowed by $\eta_{1,2}$ separately, we must have $\acomm{\eta_1}{V}=-\acomm{\eta_2}{V}\neq 0$. To show that this is not possible, we first take $V$ to be, like $\eta_j$, just a product of Pauli operators. With this assumption, the anti-commutator will be proportional to the product of Pauli operators that were in one and only one of $\eta_j$ and $V$ (or just be zero, which we are not interested in). So if $\eta_1$ and $\eta_2$ are distinct, $\acomm{\eta_1}{V}$ and $\acomm{\eta_2}{V}$ must be different by more than a minus sign and, thus, cannot add up to zero. Of course, $V$ is not generally just a product of Pauli operators, but rather a sum of products. However, this still does not allow for $\acomm{\eta_1}{V}=-\acomm{\eta_2}{V}\neq 0$ either, since the sum of two Pauli operators never yields a different Pauli operator (i.e., if $A_i$ and $B_i$ are products of Pauli operators, then $\sum_i A_i \neq \sum B_i$ unless $A_i \propto B_i$ or some permutation thereof). In other words, there is no way for the different terms in $V$ to conspire together to give some non-trivial case.}. 

Clearly, the ground states will have the same eigenvalues under $\eta$ in \equref{FirstClassOfEta} and, thus, stay real. For even system sizes,  $\eta$ in \equref{OneNaturalChoiceOfEta} are of this form and, as we have seen above, indeed admit a large class of non-Hermitian perturbations.

This is different for $\eta$ of the form of \equref{SecondClassOfEta}: the ground states will have different eigenvalues under $\eta$ and $\mathcal{P}\mathcal{T}$ symmetry is in general fragile. However, since $\eta$ in \equref{OneNaturalChoiceOfEta} can be written in the form (\ref{FirstClassOfEta}), it is clear that \equref{SecondClassOfEta} cannot be spatially homogeneous on an even-by-even lattice, but must be distinct on a non-contractible loop around the torus; the same must hold for the associated non-Hermitian perturbation, which requires, again, significant spatial fine-tuning. Let us illustrate this latter point using the concrete example of $\eta=\prod_i X_i \prod_{j\in P} Z_j$, where $P$ is a non-contractible closed path through the centers of the plaquettes. 
In that case, an imaginary field, $V=i \sum_i \sum_{\mu=1}^3 g_{i\mu} (X_i,Y_i,Z_i)_\mu$, must satisfy $g_{i1}=0$ for $i\notin P$ and $g_{i2}=0$, $g_{i1}\neq 0$ for $i\in P$ (note that $g_{i1}\neq 0$ on $P$  is required, as we otherwise can simply choose $\eta=\prod_i X_i $, which is of the form of \equref{FirstClassOfEta}, and all eigenvalues stay real). In other words, the perturbation must have vanishing $X$ components on all sites except for a non-contractible loop with non-zero $X$ components; again, not even a spatially homogeneous perturbation is possible. 

We conclude that, setting aside fine-tuned non-Hermitian perturbations with special spatial structure along non-contractible loops, suitable metric operators are of the form of \equref{FirstClassOfEta} for even-by-even lattices. As the ground states will always have eigenvalue $+1$ under any such $\eta$, the reality of their eigenvalues and, thus, $\mathcal{P}\mathcal{T}$ symmetry are protected.

\subsection{Arbitrary system sizes}
So far, we have focused our attention on even-by-even lattices since the homogeneous metric operators in \equref{OneNaturalChoiceOfEta} can be written as a product of stabilizers, while this is not possible on even-by-odd or odd-by-odd lattices; nevertheless, if one naively applies the covering shown in Fig.~\ref{fig:tcCovering} on these lattices, one can obtain a modified metric operator $\tilde{\eta}$, defined as the product of Pauli operators, $X_i$, $Y_i$, or $Z_i$, on all sites except for a single line (in the even-by-odd case) or two lines (in the odd-by-odd case) that wind around the odd lengths of the torus. In other words, $\eta$ in \equref{OneNaturalChoiceOfEta} is necessarily of the form of \equref{SecondClassOfEta} on a lattice with at least one of $L_{x}$, $L_{y}$ odd. Based on our previous discussion, this implies that the reality of the ground-state eigenvalues and $\mathcal{P}\mathcal{T}$ symmetry are generically fragile on even-by-odd and odd-by-odd  lattices.

We finally mention, for completeness, one less general but potentially useful immediate consequence. As follows from using $\tilde{\eta}$ as metric operator, any pseudo-Hermitian perturbation $\epsilon V$ with anti-Hermitian part, $\epsilon (V-V^\dagger)/2$, that has support only in a subregion of the system that is contractible around the odd lengths of the torus, will leave the ground-state eigenvalues real. 

\section{Non-Hermitian Fracton Models}\label{FractonModels}
Our analysis of the toric code carries over to many well-known fracton models in three dimensions. 
Fracton models~\cite{Chamon2005,Bravyi2011,Haah2011,VijayHaahFu1,Vijay2016, Slagle2017,Shirley2017,Ma2017,NandkishoreHermele,PretkoReview,Yoshida2013}
constitute a unique phase of matter, characterized by excitations with restricted mobility, either by being immobile or only
mobile in certain directions. These systems are typically gapped and have GSDs
exponential in linear system size.
In this section, we analyze various models with fracton order---namely, the X-cube model, checkerboard model, and Haah's codes---and show that, like the toric code, the full ground-state subspaces are stable against a large class of non-Hermitian
perturbations provided the linear system sizes along all directions are even. 
Unless stated otherwise, we take $\eta$ to be defined in the same way as in \equref{OneNaturalChoiceOfEta}, i.e., as a product of $X$, $Y$, or $Z$ operators over all qubits in the system; as motivated in Sec.~\ref{ToricCode} above in the context of the toric code, these $\eta$ provide the largest class of allowed non-Hermitian perturbations by virtue of being spatially homogeneous.

\subsection{X-cube model}

\begin{figure}
 \includegraphics[width=\linewidth]{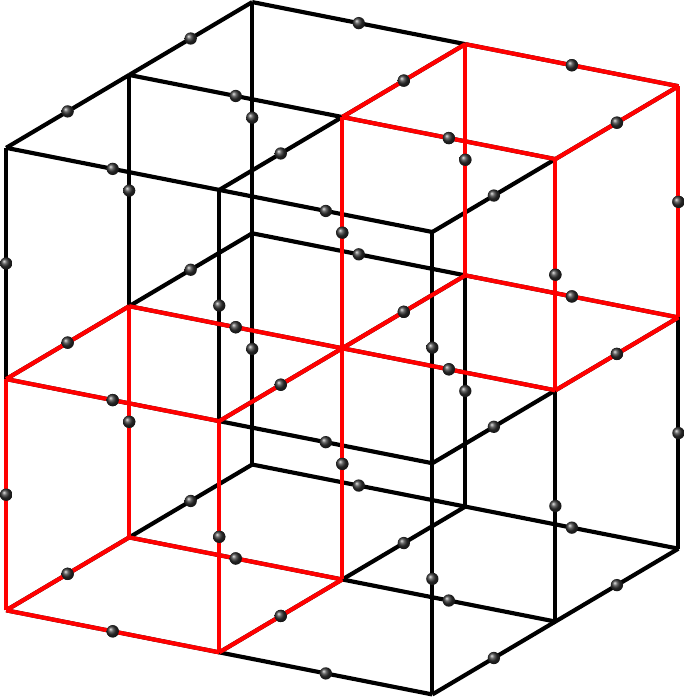}
  \caption{For the X-cube model defined on an even-by-even-by-even lattice (shown here for $2 \cross 2 \cross 2$),
    the full lattice can be covered by non-overlapping plaquette operators (red). The lattice can also be covered by vertex operators. Although this is more difficult to visualize, it can be generated by covering each 2D layer in a definite plane by a toric code covering. The remaining sites on the edges connecting these 2D layers can then be covered by chains of vertex operators.\label{fig:xcubeCovering}}
\end{figure}

The X-cube model~\cite{Vijay2016} is defined on a cubic lattice, with qubits living on the edges of the lattice.
It has a Hamiltonian composed of mutually commuting terms
\begin{equation}
  H^X = -\sum_c A_c \,\,- \hspace{-0.3em} \sum_{i = x,y,z} \sum_{v} B^i_v
  \label{eq:xcube}
\end{equation}
where $A_c = \prod_{j \in \partial c} X_j$ is the product of $X$ operators on the 12 edges of the cube labelled by $c$, and
$B^i_v$ is a vertex operator, composed of four $Z$ operators at vertex $v$ in the
plane perpendicular to the $i$'th direction.
On an even-by-even-by-even lattice, our $\eta$
operators in \equref{OneNaturalChoiceOfEta} can be assembled from these terms, thereby
showing that the entire ground-state subspace has eigenvalue
$+1$ under $\eta$, see Fig.~\ref{fig:xcubeCovering}. 
An identical argument as in the toric code case implies that $\eta$ cannot be assembled from stabilizers
on a lattice with any odd length. In combination with the fact that it commutes with all stabilizers $A_c$, $B_v^i$ separately, it must be of the form of \equref{SecondClassOfEta} for odd system lengths, with ``logical strings'' here referring to the logical string-like operators of the X-cube model.

By our analysis of the toric code, this immediately implies that the X-cube ground states on
an even-by-even-by-even lattice stay real
under the non-Hermitian perturbations permitted by $\eta$, which includes the application of imaginary
transverse fields, non-local terms like the ones considered
for the toric code, and many others. One can check that all other features of 
non-Hermitian toric code perturbations, such as their additional stability against real
perturbations and the ability to add contractible perturbations on lattices with odd
system sizes, also hold. However, these features are more striking for fracton models: instead of a four-dimensional code subspace being protected against these perturbations,
fracton models have a GSD that grows exponentially with system size; for the X-cube model on a three-dimensional torus, the
GSD obeys 
\begin{equation*}
  \begin{aligned}
    \log_2 \text{GSD} = 2 L_x + 2 L_y + 2 L_z - 3.
  \end{aligned}
\end{equation*}
The reality of the code subspace in the presence of pseudo-Hermitian perturbations holds for the X-cube model defined on general three-dimensional manifolds~\cite{Shirley2017}, provided
the full space can be covered by plaquette or star operators.

This sensitivity to system size may be surprising, since the X-cube model
exhibits \textit{foliated fracton order}~\cite{Shirley2017}. This means that
the length of any of the sides of the X-cube model can always be extended 
by attaching layers of toric code and applying a series of local unitary
transformations. In Appendix~\ref{ap:foliation}, we present a detailed study
of how the metric operators $\eta$ behave under foliations. The end
result is that, while the ground states can be extended by this foliation procedure,
the foliation acts non-trivially on $\eta$, meaning that the interplay between
$\eta$ and the X-cube ground states can change depending on the system size.

\subsection{Checkerboard model}

The checkerboard model~\cite{Vijay2016} is another example of a system with fracton excitations. This model has spins defined on the vertices
of a three-dimensional cubic lattice, as opposed to the edges. By separating the cubes of the lattice with alternating
labels $A$ and $B$, each forming a three-dimensional checkerboard lattice, and denoting the cubic operators $\prod_{i \in \partial c} Z_i$ and $\prod_{i \in \partial c} X_i$ as $Z_c$ and $X_c$ respectively, 
the checkerboard model is given by the Hamiltonian 
\begin{equation}
  H^{C} = -\sum_{c \in A} Z_c - \sum_{c \in A} X_c.
  \label{eq:checkerboard}
\end{equation}
The geometry of the checkerboard model requires it to be defined
on an even-by-even-by-even lattice if periodic boundary conditions are imposed, since otherwise one cannot uniformly
partition the cubes into $A$ and $B$ labels.
On even-by-even-by-even lattices, the entire lattice can be covered by non-overlapping stabilizers, and therefore
the ground-state subspace is even under any $\eta$ in \equref{OneNaturalChoiceOfEta}. Again, this implies that the ground-state energies of the 
checkerboard model always
remain real under non-Hermitian perturbations that are pseudo-Hermitian under $\eta$. Although only small system sizes are accessible via ED, we have checked these predictions numerically for the checkerboard model on a $2 \cross 2\cross 2$ lattice. 

A Majorana version of the checkerboard model has also been studied~\cite{VijayHaahFu1}, 
which simply replaces the Pauli spins with Majorana fermions $\gamma_i$, i.e., the model has one Majorana fermion per site $i$ of the cubic lattice. By defining
$\prod_{i \in c} \gamma_i = \gamma_c$, the Hamiltonian of the Majorana checkerboard model is
\begin{equation}
  \begin{aligned}
    H = -\sum_{c \in A} \gamma_c. \label{MajoranaCheckerboard}
  \end{aligned}
\end{equation}
Because the entire lattice can be covered with $\gamma_{c \in A}$,
all ground states of the system are even under the operator
\begin{equation*}
  \begin{aligned}
    \eta = \prod_i \gamma_i\,,
  \end{aligned}
\end{equation*}
which can be interpreted as the total fermion parity, $\eta \propto \prod_{\alpha} (c^\dagger_\alpha c^{\phantom{\dagger}}_\alpha - 1/2)$, when combining pairs of Majorana fermions into auxiliary complex fermions $c_\alpha$.
Therefore, the ground states remain real under
perturbations of the form $i\epsilon\mathcal{O}$, where each term in $\mathcal{O}$ contains an odd number of Majorana
operators, i.e., changes the total occupation of auxiliary complex fermions by an odd amount.

\subsection{Haah's codes}

Finally, we consider Haah's 17 CSS cubic codes~\cite{Haah2011}, all of which are defined on a cubic lattice with two qubits 
per site $i$. Each cube has two stabilizers: one is built up of tensor products of $Z$ and $\mathbbm{1}$ 
operators on each site $i$, such as $Z_i\otimes \mathbbm{1}_i$ or $Z_i\otimes Z_i$; the other one involves 
tensor products of $X$ and identity operators, e.g., $X_i\otimes \mathbbm{1}_i$.
The exact form of the stabilizers differs from code to code, but all have a sub-extensive GSD. We defer a 
more detailed discussion of these codes to Appendix~\ref{ap:haah}---our conclusion is that, with the choice 
of $\eta$ analogous to \equref{OneNaturalChoiceOfEta},
\begin{equation}
    \eta = \prod_i X_i \otimes X_i, \, \prod_i Y_i \otimes Y_i,\, \prod_i Z_i \otimes Z_i \,, \label{EtasForTheHaahCodes}
\end{equation}
the behavior of the code subspace under pseudo-Hermitian perturbations is sensitive not only to whether
the system lengths are even or odd, but also whether the system lengths are divisible by $4$. Moreover,
since not all cubic codes are symmetric under rotations, this behavior is dependent on which directions
are even or odd, and which are divisible by $4$. This admits eight different classes of codes, based on
the relation between their code subspace stability under pseudo-Hermitian perturbations and their system sizes.
These classes range from cubic code $7$, whose code subspace stays real on all system sizes other than
odd-by-odd-by-odd, and cubic code $17$, where the code subspace only stays real if $L_x$, $L_y$ are divisible
by $4$. We refer to Appendix~\ref{ap:haah} for a complete characterization of this behavior. 

\section{Non-Hermitian Quantum Fractal Liquids}\label{FractalLiquids}
In this section, we will generalize the previous analysis to also include another class of fracton models dubbed ``quantum fractal liquids'' \cite{Yoshida2013} and reformulate the criterion of stability against non-Hermitian perturbations using a polynomial representation of Pauli operators. In this way, we will recover the criterion of stability of the toric code in an algebraic way and show that the reality of eigenvalues of the exponentially large number of ground states of quantum fractal liquids is protected against a wide range of non-Hermitian terms in the Hamiltonian.

\subsection{Polynomial representation of operators}\label{PolynomialRepresentation}
To set up the notation, we will briefly introduce the polynomial representation of operators, a commonly used technique \cite{macwilliams1977theory}. To this end, consider a polynomial of three variables, $x$, $y$, $z$, 
\begin{equation*}
  \begin{aligned}
    f = \sum_{j, k, \ell \in \mathbbm{Z}} c_{jkl}\, x^j y^k z^\ell,\quad c_{jk\ell} = 0, 1,
  \end{aligned}
\end{equation*}
over $\mathbb{F}_2$, meaning that all coefficients are to be understood modulo $2$.
This allows to define a corresponding Pauli operator whose components lie on the vertices of a cubic lattice in three dimensions in the following way
\begin{equation*}
  \begin{aligned}
    Z(f): = \prod_{jk\ell} Z_{jk\ell}^{c_{jk\ell}}, \quad X(f) := \prod_{jk\ell} X_{jk\ell} ^{c_{jk\ell}}\,.
  \end{aligned}
\end{equation*}
Here $Z_{jk\ell}$ ($X_{jk\ell}$) is the $Z$ ($X$) operator acting at vertex $(j,k,\ell)$.
For example, a stabilizer of the checkerboard model,
given by the product of Pauli matrices on the eight vertices
of a cube, corresponds to the polynomial $f = 1 + x + y + z + xy + yz + xz + xyz$.
On a finite lattice, periodic boundary conditions are specified by imposing $x^{L_x} = y^{L_y} = z^{L_z} = 1$. We denote the dual of $f$, obtained by taking $x \rightarrow x^{-1}$, and likewise for $y$ 
and $z$, by $\bar{f}$.

Certain relations can be expressed more concisely with this polynomial representation. Translating an operator $Z(f)$ one lattice site along the $x$-direction is simply given by $Z(xf)$, and likewise for translations in the $y$ and $z$-direction.
Additionally, the polynomials defined over $\mathbb{F}_2$
naturally encode the commutation relations of the Pauli operators. To see this, consider the \textit{commutation polynomial}, defined as $f\bar{g}$ for two polynomials $f$ and $g$. Writing $f \bar{g}$ as
\begin{equation*}
  \begin{aligned}
    f \bar{g} = \sum_{ijk} d_{ijk}\, x^i y^j z^k\,,
  \end{aligned}
\end{equation*}
$d_{ijk} = 1$ ($0$) implies that $Z(f)$ and $X(x^i y^j z^k g)$ anti-commute (commute).

Quantum fractal liquids are defined on a cubic lattice with two spins on every vertex. The form of their stabilizers is given by \cite{Yoshida2013}
\begin{align}\begin{split}
    Z(\alpha, \beta),& \quad X(\bar{\beta},\bar{\alpha}), \\
    \alpha = 1-f(x) y,& \quad \beta =1-g(x) z,
    \label{eq:fractalStabilizers}
\end{split}\end{align}

and translations thereof, where the two arguments of $Z$ and $X$ denote operators on the two distinct spins per site. Different choices of polynomials $f$ and $g$ define different models. Clearly, all stabilizers commute, as follows from the associated commutation polynomial, $\alpha\bar{\bar{\beta}} + \beta \bar{\bar{\alpha}} = 2\alpha\beta =0$. 

For codes defined by stabilizers of this type, the logical operators take the form
\begin{equation}\begin{split}
  \begin{aligned}
    &\ell_i^{(Z)} = Z(0, x^i \bm{f}(x,y))\,,\quad r_i^{(Z)} = Z(x^i \bm{g}(x, z), 0) \,,
    \\
    &\ell_i^{(X)} = X(x^i \bar{\bm{f}}(x,y), 0) \,,\quad r_i^{(X)} = X(0, x^i \bar{\bm{g}}(x,z)),
  \end{aligned}\label{StringOperators}\end{split}
\end{equation}
for integer $i=0,1,\dots, L_x-1$, where we define
\begin{equation}
    \bm{f} = \sum_{k=1}^{L_y} (fy)^{k-1}, \quad \bm{g} = \sum_{\ell=1}^{L_z} (gz)^{\ell-1}.
\end{equation}
It is straightforward to verify that the operators in \equref{StringOperators} commute with the stabilizers and constitute logical operators if
\begin{equation}
    f^{L_y} = 1, \quad g^{L_z} = 1. \label{ConditionForStabilizers}
\end{equation}
There are various ways to satisfy \equref{ConditionForStabilizers}: the ``trivial'' solution, that works for any set, $L_x$, $L_y$, $L_z$, of system sizes, is $f=g=1$. This corresponds to layers of toric code in the ($\hat{y}$,$\hat{z}$) plane, upon noting that the bond variables of the toric code, see \figref{fig:tcCovering}, can be seen as two qubits per vertex. In this case, $\ell_i^{(Z,X)}$ and  $r_i^{(Z,X)}$ in \equref{StringOperators} become Z-, X-type string operators in the $i$th layer along the $\hat{y}$ and $\hat{z}$ direction, respectively. Another way of satisfying \equref{ConditionForStabilizers} that works for arbitrary isotropic system sizes, $L_x=L_y=L_z=L$, is $f=x^{n_f}$, $g=x^{n_g}$. However, the largest class of possible polynomials $f$, $g$ and, thus, possible models is allowed in the isotropic case with $L=2^{n_L}$, since \equref{ConditionForStabilizers} will hold as long as $f(1)=g(1)=1$ \cite{Yoshida2013}. Here, we refer to the latter set of models as ``quantum fractal liquids,'' which have been shown to exhibit exponential scaling of the GSD, obeying $ \log_2 \text{GSD}(2L) = 2 \log_2 \text{GSD}(L)$ \cite{Yoshida2013}.
Note, however, that the absence of string-like logical operators and mobile quasiparticles further requires that $f$ and $g$ are not algebraically related, i.e., that there are no integers $n_1$ and $n_2$ such that $f^{n_1} = g^{n_2}$ (neglecting periodic boundary conditions). An example of a model free of string-like logical operators is provided by $f=1+x+x^2$ and $g=1+x+x^3$.

\subsection{Pseudo-Hermitian perturbations}
As before, we are interested in adding pseudo-Hermitian perturbations to this class of models that will leave the ground-state subspace real. We take $\eta$ to be defined analogous to \equref{EtasForTheHaahCodes} or, in polynomial representation, 
\begin{align}\begin{split}
    \eta &= Z(h,h), \, X(h,h), \, iX(h,h)Z(h,h), \\
    h & = \sum_{j=1}^{L_x}\sum_{k=1}^{L_y}\sum_{\ell=1}^{L_z} x^{j-1}y^{k-1}z^{\ell-1}.
\label{EtaExpressedByPolynomials}\end{split}\end{align}
Any $\eta$ in \equref{EtaExpressedByPolynomials} will commute with all stabilizers (\ref{eq:fractalStabilizers}). This readily follows from the associated commutation polynomial upon noting that $h=\bar{h}$ is invariant under multiplication by any monomial, physically related to the translation invariance of $\eta$, and that the number of monomials in both $f$ and $g$ must be odd. The latter is a consequence of \equref{ConditionForStabilizers} and of the observation that the parity of the number of terms of a polynomial $f$ over $\mathbb{F}_2$ is the same as that of any of its powers, $f^n$ with $n>0$.

Based on our discussion of Sec.~\ref{GeneralFormOfPerturbations}, we want to analyze under which conditions the ground-state subspace is even under these operators to guarantee that their eigenvalues stay real. Previously, we had verified this by
attempting to assemble $\eta$ via the stabilizers of the model. In the set of models introduced above, the polynomial
representation makes it easier to instead verify whether $\eta$ commutes with all the logical operators (\ref{StringOperators}), which in turn implies that all ground states have the same eigenvalue of $\eta$ [and that $\eta$ is of the form of \equref{FirstClassOfEta} rather than \equref{SecondClassOfEta}].

The condition for $\eta$
to commute with all the logical string operators is given by
\begin{equation}
  \begin{aligned}
   & h \bm{g} = h \bm{f}  = 0,
   \label{eq:fractalCommutation}
  \end{aligned}
\end{equation}
for any of the three possible choices in \equref{EtaExpressedByPolynomials}.
This simple expression arises from the fact that $\bar{h}=h$ and that the logical operators come in exactly the form of operators relevant to the commutation polynomial, so one can 
verify that $\eta$ commutes with all the string operators
with one equation. It would technically suffice for $h \bm{g}$
and $h \bm{f}$ to be only a function of $y$ and $z$, since
one is only concerned with the commutations of operators like 
$Z(h)$ and $X(x^i \bm{f})$, but not those with relative shift along the $\hat{y}$ or $\hat{z}$ directions. However, recalling that $h$ is invariant under
multiplication by any monomial, there is no way for
$\bm{g}$ or $\bm{f}$ to conspire to cancel out only the terms independent of $y$ and $z$ in $h \bm{g}$ and $h \bm{f}$ without simply giving $0$.

Another important consequence of $h$ being invariant under the multiplication by any monomial is that \equref{eq:fractalCommutation} is satisfied if and only if
$\bm{f}$ and $\bm{g}$ contain an even number of monomials. As argued above, \equref{ConditionForStabilizers} implies that $f$, $g$ and, therefore, also $f^n$, $g^n$ must contain an odd number of terms. Taken together, \equref{eq:fractalCommutation} is obeyed and, thus, \textit{the reality of the eigenvalues of the ground states is protected against pseudo-Hermitian perturbation with $\eta$ given in \equref{EtaExpressedByPolynomials} if $L_y$ and $L_z$ are even.} Note that the $x$-direction is distinguished 
from the other two directions in this criterion, a reflection
of the fact that the stabilizers given by \equref{eq:fractalStabilizers} also distinguishes the $x$-direction.

Let us illustrate this for the different special cases of $f$ and $g$ noted above. Taking $f=g=1$ corresponds to $L_x$ uncoupled layers of toric code and the above statement implies that the toric code is protected if and only if the number of sites in each in-plane direction is even, reproducing the result of Sec.~\ref{ToricCode}. Our current formalism, however, captures many more cases. For instance, we immediately conclude that any model with $f=x^{n_f}$, $g=x^{n_g}$ and $L_x=L_y=L_z=L$ is protected only for even $L$. As the two polynomials are algebraically related, this two-parameter family of models is characterized by string-like logical operators and has excitations mobile along the direction $n_g \hat{y} - n_f \hat{z}$ \cite{Yoshida2013}. Finally, as already noted above, quantum fractal liquids with arbitrary $f$ and $g$, only constrained by $f(1)=g(1)=1$, are in general defined on lattices with an even number of sites and, as such, are always protected against pseudo-Hermitian perturbations with metric operator in \equref{EtaExpressedByPolynomials}.

\section{Summary and Conclusions}\label{Conclusion}
In this work, we studied the behavior of the eigenvalues of quantum many-body Hamiltonians of the form of \equref{CompleteDesciptionOfHam}, i.e., starting from a Hermitian system, $H_0$, we turn on a non-Hermitian perturbation, $\epsilon V$, and demand that the entire Hamiltonian be pseudo-Hermitian. Using pseudo-Hermiticity rather than $\mathcal{P}\mathcal{T}$ symmetry is related to the fact that the former is more general than the latter \cite{2018arXiv180101676Z}; we note, however, that all of the explicit examples considered here are both $\mathcal{P}\mathcal{T}$ symmetric and pseudo-Hermitian. We analyzed whether the energies, $E_i$, of a given subspace of interest of $H_0$ will remain real as long as the gap to other states of the system is finite ($\mathcal{P}\mathcal{T}$ symmetry protected) or whether they can move into the complex plane without closing the gap ($\mathcal{P}\mathcal{T}$ symmetry fragile). 
While symmetries can enforce degeneracies ($E_i = E_{i'}$) and protect eigenvalues from becoming complex in conjunction with pseudo-Hermiticity ($E_i=E_{i'}^*$), we discussed that this is also possible in the absence of symmetries: if the eigenvalues of the metric operator $\eta$ are the same for all states in the subspace of interest, $E_i$ are guaranteed to stay real and $\mathcal{P}\mathcal{T}$ symmetry is protected. 

We demonstrated that this criterion can be readily applied to various paradigmatic many-body models with crucial implications. As a first example, we took the toric code model (\ref{eq:toricCode}) as unperturbed Hamiltonian, $H_0$. On a torus, it exhibits four degenerate ground states and one would generically expect them to become complex when turning on $\epsilon V$. However, we have shown that $\eta$ of the form given in \equref{OneNaturalChoiceOfEta} allows for a large class of non-Hermitian perturbations; these are shown to leave the ground-state energies real on an even-by-even lattice, even if all symmetries are broken. They can only become complex and $\mathcal{P}\mathcal{T}$ can only be broken in the ground-state subspace, when the gap to the excited states closes. In fact, we have argued that any sufficiently generic non-Hermitian perturbation (see Sec.~\ref{OtherMetricOperators}) in a system with both linear system sizes even (at least one of them odd) will only allow for $\eta$ of the form of \equref{FirstClassOfEta} [of the form of \equref{SecondClassOfEta}] and the ground-state eigenvalues are protected (not protected) from becoming complex. This sensitivity to system size reflects the highly entangled nature of the toric-code ground states.

We came to the same conclusions for the ground-state manifolds of the X-cube (\ref{eq:xcube}), the spin (\ref{eq:checkerboard}) and Majorana (\ref{MajoranaCheckerboard}) checkerboard models, and for the fractal liquids of Ref.~\cite{Yoshida2013}. In these cases, the stability of $\mathcal{P}\mathcal{T}$ symmetry is even more surprising due to the enormous GSD that grows exponentially with system size. For Haah's 17 codes, the stabilizers have a slightly more complicated form and the minimal requirement for stability differs from code to code, although we observe several groups of codes which all obey the same requirements. This classification of Haah's codes based on stability of $\mathcal{P}\mathcal{T}$ symmetry approximately follows previous classifications based on entanglement renormalization~\cite{Dua2019}.

On a more general level, our work illustrates that $\mathcal{P}\mathcal{T}$ symmetry and the reality of energies can be protected in the degenerate ground-state manifold of correlated many-body systems with different forms of topological order---even in the absence of any symmetries and although exceptional points are generically expected to be ambundant \cite{PiazzaQMB}.
By virtue of being exact and simple, our framework can be readily applied to a large class of systems and provides a systematic method for constructing pseudo-Hermitian perturbations that ensures the reality of the resulting eigenvalues. This is not only relevant for experimental studies \cite{Jorg2019, Takasu2020,Regensburger2012,LeiPTQuantumDynamics} and potential applications \cite{TopologicalLaser1,TopologicalLaser2,TopologicalLaser3,WiersigSensing,LiuSensing}, but might also help deepen our theoretical understanding of non-Hermitian systems hosting exotic phases of matter, e.g., by providing novel ways of classifying spin-liquid or fracton phases according to their sensitivity to such perturbations.

\section*{Acknowledgments}

This research was supported by the National Science Foundation under Grant No.~DMR-1664842.  
We thank Arpit Dua, Darshan Joshi, Saranesh Prembabu, Subir Sachdev, Robert-Jan Slager, and Nathan Tantivasadakarn for helpful discussions.

\appendix
\section{Perturbative derivation of the condition for reality of eigenvalues}
\label{ap:perturbation}
Here we provide a formal derivation of the statement of Sec.~\ref{GeneralFormOfPerturbations} of the main text that the eigenvalues of any (almost) degenerate subspace of $H_0$ in \equref{GeneralFormOfHam} will remain real upon adiabatically turning on the non-Hermitian perturbation $\epsilon V$, if all states in the (almost) degenerate subspace have the same eigenvalue under the metric operator $\eta$. 
We will discuss two different perturbative expansions and prove that the above holds true to all orders. We will then discuss the approximate orthogonality of the associated eigenstates.

To this end, we will consider a pseudo-Hermitian Hamiltonian of the form of \equref{GeneralFormOfHam},
\begin{equation}
    H_{\epsilon} = H_0 + \epsilon \, V, \quad \epsilon\in\mathbbm{R}, \label{FullHamiltonian}
\end{equation}
and a metric operator $\eta$, such that $\comm{\eta}{H_0} = 0$ for the Hermitian unperturbed part, $H_0^{\pdagger}=H_0^\dagger$,
and $\eta V \eta^{-1} = V^\dagger$ for the perturbation.

We are interested in the behavior of the eigenvalues of a subset of (orthonormal) eigenstates, $\{\ket{\psi_i},i=1,\dots,n\}$, of $H_0$, which can be arbitrarily close or identical in energy but are well separated from all other eigenvalues. We refer to the space spanned by $\{\ket{\psi_i},i=1,\dots,n\}$ as the almost degenerate subspace.

To analyze how their eigenvalues, $E_i(\epsilon)$, $i=1,2,\dots, n$, evolve when turning on $\epsilon V$ in \equref{FullHamiltonian}, we define the projectors $P$ and $Q$,
\begin{equation*}
    P =  \sum_{i=1}^n \ket{\psi_i}\bra{\psi_i}, \quad Q = \mathbbm{1} - P\,,
\end{equation*}
to the almost degenerate subspace and its complement. We use that the exact eigenstates, $\ket{\Psi_i(\epsilon)}$, obeying
\begin{equation*}
    H_\epsilon \ket{\Psi_i(\epsilon)} = E_i(\epsilon) \ket{\Psi_i(\epsilon)},
\end{equation*}
must also satisfy \cite{Buth2004}
\begin{equation}
    H^{\text{eff}}_\epsilon(E_i(\epsilon)) P \ket{\Psi_i(\epsilon)} = E_i(\epsilon) P \ket{\Psi_i(\epsilon)} \label{EffectiveSchrEq}
\end{equation}
with the effective Hamiltonian 
\begin{align}
    H^{\text{eff}}_\epsilon(E) &= P H_{\epsilon} P + P H_{\epsilon} Q G_{\epsilon}(E) Q H_{\epsilon} P, \label{EffectiveHam}
    \\
    G_{\epsilon}(E) &= \left[ E - Q H_{\epsilon} Q \right]^{-1}. \label{GreensFunction}
\end{align}
As follows from \equref{EffectiveSchrEq}, the eigenvalues $E_i(\epsilon)$, $i=1,2,\dots, n$, can be obtained by diagonalizing the effective Hamiltonian $H^{\text{eff}}_\epsilon$ in the almost degenerate subspace. Of course, this is not straightforward to do as the effective Hamiltonian itself depends on these eigenvalues; however, the effective-Hamiltonian formulation is a good starting point to develop a perturbative expansion. 

\subsection{Expansion in $\epsilon$}
Since we view the non-Hermitian part $\epsilon V$ as a perturbation to $H_0$ in our discussion in the main text, it is very natural to expand in $\epsilon$. Note that $P H_0 Q = 0$, so the 
second term in the effective Hamiltonian (\ref{EffectiveHam}) is $\order{\epsilon^2}$,
\begin{equation}
    H^{\text{eff}}_\epsilon(E) = P H_0 P + \epsilon P V P + \epsilon^2 P V Q G_{\epsilon}(E) Q V P. \label{SimplifiedEffHam}
\end{equation}
Let us now assume that we can obtain $E_i(\epsilon)$ via perturbative expansion in $\epsilon$. To keep the notation compact, let us define the operator $T_N$ which performs a Taylor expansion on a function or operator up to and including order $N$, i.e., $T_N[f(x)] := \sum_{k=0}^N \frac{x^k}{k!} \frac{\textrm{d} f}{\textrm{d} x}(0)$. As follows from \equref{SimplifiedEffHam}, $T_1[E_i(\epsilon)]$, for any $i=1,2,\dots , n$, is obtained by diagonalization of
\begin{equation}
    h^{(1)}_{ij} := \bra{\psi_i}( H_0 + \epsilon V ) \ket{\psi_j}, \quad i,j=1,2,\dots , n. \label{LeadingOrderEffHam}
\end{equation}
Since, by construction, all $\ket{\psi_i}$ have the same eigenvalue under $\eta$, we conclude from \equref{HermitianMatrixInEigenspace} that $h^{(1)}_{ij}$ is Hermitian and, thus, $T_1[E_i(\epsilon)] \in \mathbbm{R}$. Higher orders, $T_{N >1}[E_i(\epsilon)]$, are obtained by iteratively diagonalizing
\begin{align}\begin{split}
    h^{(N)}_{ij} &:= \bra{\psi_i}( H_0 + \epsilon V \\ &\quad + \epsilon^2 T_{N-2}[ V Q G_{\epsilon}(T_{N-2}[E_i(\epsilon)]) Q V ] ) \ket{\psi_j}. \label{NthOrderDiagonalization}
\end{split}\end{align}
First, note that $Q$ commutes with $\eta$ which implies that $G_{\epsilon}(E)$ and, thus, also $V Q G_{\epsilon}(E) Q V$ are pseudo-Hermitian if $E\in\mathbbm{R}$. Since this holds for a continuous set of values of $\epsilon$, this property holds for each order in the Taylor expansion separately. As such, it also applies to $T_{N-2}[V Q G_{\epsilon}(E) Q V]$ in \equref{NthOrderDiagonalization}. Due to $T_1[E_i(\epsilon)] \in \mathbbm{R}$, iterative diagonalization of \equref{NthOrderDiagonalization} will always yield real eigenvalues. Taken together we have shown that $E_i(\epsilon)$, $i=1,2,\dots, n$, stay real to any order in $\epsilon$. 

If the eigenstates are exactly degenerate for $\epsilon=0$, the leading non-zero contribution to the energy splitting will determine whether the eigenvalues of $H_{\epsilon}$ stay real or become complex. In most cases, the first order corrections,
given by diagonalizing $P H_\epsilon  P$, break the degeneracy. Since $P H_\epsilon  P$
is clearly Hermitian, our result is simple if the first order energy splitting is non-zero. In fact, a mathematical proof to first order in perturbation theory has been provided in Ref.~\cite{Caliceti2004}. However,
topological degeneracies are often broken only at higher orders in perturbation theory, so a more general result
is required.

If, however, the degeneracy is already broken for $\epsilon=0$, our current perturbative approach cannot be used to understand whether the eigenvalues stay real or not: by construction, we assume that $E_{i}(\epsilon)$ is an analytic function of $\epsilon$ and therefore will never be able to reproduce the $\epsilon$ dependence of real eigenvalues meeting and moving into the complex plane. For this reason, we next present an alternative approach.

\subsection{Expansion in energy separation}
The problem noted above that arises when the eigenstates of $H_0$ are not exactly degenerate can be reconciled by performing an expansion in the energy gap between the almost degenerate subspace and the rest of spectrum. More formally, we generalize the effective Schr\"odinger equation (\ref{EffectiveSchrEq}) by introduction of a dimensionless parameter $\lambda$,
\begin{equation}
    H^{\text{eff}}_{\epsilon,\lambda}(E_{i\epsilon}(\lambda)) P \ket{\Psi_{i\epsilon}(\lambda)} = E_{i\epsilon}(\lambda) P \ket{\Psi_{i\epsilon}(\lambda)},
\end{equation}
where 
\begin{equation}
    H^{\text{eff}}_{\epsilon,\lambda}(E) = P H_{\epsilon} P + \lambda\, P H_{\epsilon} Q G_{\epsilon}(E) Q H_{\epsilon} P. \label{ModifiedEffectiveHam}
\end{equation}
We assume that we can expand $E_{i\epsilon}(\lambda)$ in a power series of $\lambda$, but treat its $\epsilon$-dependence exactly, and show that it stays real to all orders in $\lambda$. Since $\lambda$ multiplies $G_{\epsilon}$ in \equref{ModifiedEffectiveHam}, this expansion is controlled by the gap to the other states of the spectrum being large (compared to $\epsilon V$). The argument proceeds similar to the one above: the zeroth order contribution, $T_0[E_{i\epsilon}(\lambda)] = E_{i\epsilon}(0)$, is obtained from diagonalization of \equref{LeadingOrderEffHam} and as such real. One can compute $T_N[E_{i\epsilon}(\lambda)]$ from $T_{N-1}[E_{i\epsilon}(\lambda)]$ by iterative diagonalization of 
\begin{align*}\begin{split}
    &h^{(N)}_{ij} := \\ &\,\,\,\bra{\psi_i}( H_\epsilon + \lambda\, T_{N-1}[ H_\epsilon Q G_{\epsilon}(T_{N-1}[E_{i\epsilon}(\lambda)]) Q H_\epsilon ] ) \ket{\psi_j}.
\end{split}\end{align*}
With the same arguments as above, this implies that $T_N[E_{i\epsilon}(\lambda)]$ will stay real for any $N$.
Of course, the perturbative approach is expected to break down when the gap between the almost degenerate subspace and another part of the spectrum with different eigenvalue under $\eta$ closes since $G_{\epsilon}$ will develop a pole.

\subsection{Approximate orthogonality}
Above, we have argued that the effective Hamiltonians in Eqs.~(\ref{EffectiveHam}) and (\ref{ModifiedEffectiveHam}) will be Hermitian if the eigenvalues of $\eta$ are identical in the almost degenerate subspace. This does not only have consequences for the reality of the eigenvalues, but also for their mutual orthogonality. 

To first order in $\epsilon$ and zeroth order in $\lambda$, i.e., to leading order in the limit of a large gap to the excited states, the effective Hamiltonian is also independent of $E$. Therefore, the projections $P\ket{\Psi_i(\epsilon)}$, $i=1,2,\dots n$, are obtained as eigenstates of the same Hermitian Hamiltonian and, as such, orthogonal. 
Naturally, this does not mean that $\ket{\Psi_i(\epsilon)}$ are orthogonal in the full Hilbert space; however, the differences between the full and the projected states, $\ket{\Psi_i(\epsilon)}-P\ket{\Psi_i(\epsilon)} = Q\ket{\Psi_i(\epsilon)}$, are also suppressed in the limit of large energetic separation to the rest of the spectrum since \cite{Buth2004}
\begin{equation}
    Q\ket{\Psi_i(\epsilon)} = \epsilon G_{\epsilon}(E_{i}(\epsilon))QVP\ket{\Psi_i(\epsilon)},
\end{equation}
as stated in the main text.

\section{Interplay between X-Cube foliation and metric operators}
\label{ap:foliation}
In the main paper, we noted that the ground states of the X-cube model all
have the same eigenvalue under our choice of metric operator $\eta$ in \equref{OneNaturalChoiceOfEta}, provided all
lengths are even. This is because $\eta$ can be assembled by
a collection of stabilizers. While $\eta$ cannot be assembled
by stabilizers on a system with odd lengths, it is known that the X-cube model exhibits 
\textit{foliated fracton order}~\cite{Shirley2017}, which implies that
an $L \cross L \cross L$ X-cube model ground state can be enlarged to a ground state of an
$L \cross L \cross L+1$ model by the attachment of an $L \cross L$ toric
code ground state and the application of local unitary operators. If $L$ is
even, then the original X-cube ground states and the toric code ground states
will all have the same eigenvalue under $\eta$. Because of this, one may suspect
that the resulting $L \cross L \cross L + 1$ ground states may also have the same
eigenvalue under the appropriately enlarged $\eta$. However, as we will show,
the process of attaching the two states and applying local unitary operators
yields an $L \cross L \cross L + 1$ state that is not an eigenstate
of the enlarged $\eta$.

\begin{figure}
\includegraphics[width=0.8\linewidth]{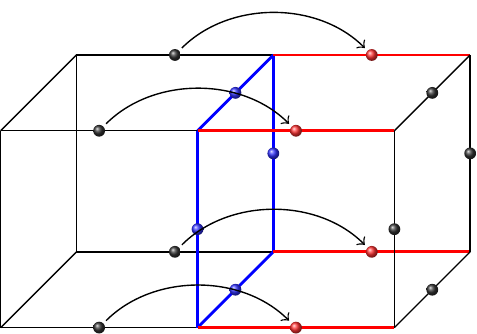}
\includegraphics[width=0.8\linewidth]{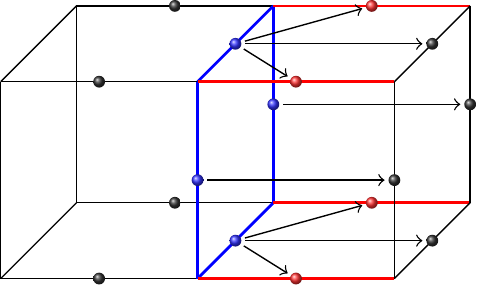}
    \caption{The size of an X-cube model ground state can be increased along
    one axis by adding a layer of toric code (blue) and an additional set of
    qubits initialized in the $\ket{0}$ state (red). A series of CNOT gates are
    applied to this tensor product of states to yield an enlarged X-cube model
    ground state. The application of the CNOT gates is shown above,
    with arrows pointing from the control to the target qubit. The CNOT gates
    are applied in two steps---the gates in the top diagram are used first,    then the gates in the bottom diagram are applied.}
    \label{fig:foliation}
\end{figure}

We first describe the process of adding an extra layer to the X-cube model, 
illustrated in Fig.~\ref{fig:foliation}. We begin with an $L \cross L \cross L$
X-cube ground state, $\ket{\psi_X}$, an $L \cross L$ toric code ground state,
$\ket{\psi_{TC}}$, and a collection of $L^2$ additional qubits initialized in
the $\ket{0}$ state, $Z \ket{0} = \ket{0}$. The statement of foliated fracton order
is that an $L \cross L \cross L + 1$ X-cube ground state, $\ket{\psi'_X}$, can
be written as
\begin{equation*}
    \ket{\psi'_X} = S \left( \ket{\psi_X} \otimes \ket{\psi_{TC}} \otimes \ket{0}^{L^2} \right)
\end{equation*}
where $S$ is a series of local unitary transformations, which in our case is
given by a collection of CNOT gates \cite{Shirley2017}. 

\begin{figure}
\includegraphics[width=0.8\linewidth]{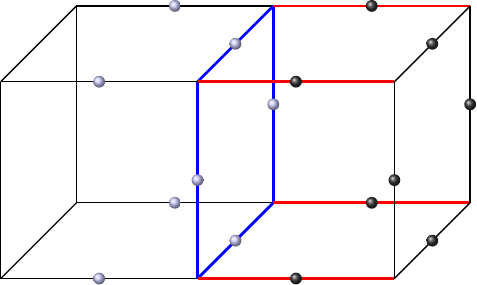}
\includegraphics[width=0.8\linewidth]{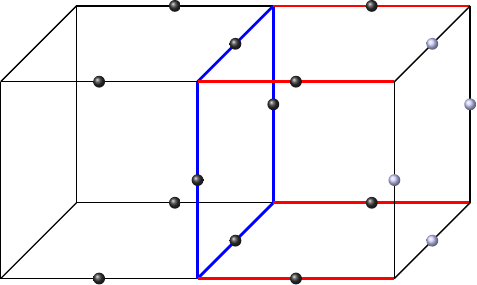}
    \caption{The behavior of the metric operator $\eta$ on an $L \cross L \cross L + 1$ X-cube model ground state can be calculated by an effective operator, $S^\dagger \eta S$, acting on the exfoliated parts of the X-cube model. Here, we
    show the action of $S^\dagger \eta_Z S$ (top) and $S^\dagger \eta_X S$ (bottom), where the effective operator is the product of Z (X) operators on all the dark sites. The tensor product of the exfoliated parts of the X-cube ground state is
    not generally an eigenstate of the corresponding effective operators. The decomposition of $\eta_Y$ is identical to that of $\eta_X$.}
    \label{fig:foliationMetricOperator}
\end{figure}

This foliation allows us to deduce the behavior of $\eta$ in \equref{OneNaturalChoiceOfEta} applied to $\ket{\psi'_X}$
based on the action of $S^\dagger \eta S$ on the three constituent states,
assuming $L$ is even.
This behavior is dependent on the form of $\eta$. We first begin with an 
analysis of $\eta_Z \equiv \prod_i Z_i$. Carrying out the corresponding
CNOT gate transformations, we see in Fig.~\ref{fig:foliationMetricOperator} that
the action of $S^\dagger \eta_Z S$ on the original X-cube ground state is not simply
the product of all $Z_i$ operators---some sites are missing in a way that cannot simply be compensated by a product of stabilizers; this means that
$\ket{\psi_X} \otimes \ket{\psi_{TC}} \otimes \ket{0}^{L^2}$ will generally not be an eigenstate of $S^\dagger \eta_Z S$. Carrying this through with $\eta_X = \prod_i X_i$ and $\eta_Y = \prod_i Y_i$ yields
a similar result. In accordance with the analysis of the main text, we conclude that not every ground state of an even-by-even-by-odd X-cube model
 will be an eigenstate of $\eta$, as the foliation 
process complicates the behavior of the metric operator. 

One can take this
$L \cross L \cross L + 1$ model and attach additional toric code layers
in either of the two remaining directions, and an identical analysis implies
that even-by-odd-by-odd and odd-by-odd-by-odd ground states will not all have
the same eigenvalue under $\eta$. Of course, one can add another toric code layer
to give an $L \cross L \cross L + 2$ model, in which case the metric operator 
\textit{does} decompose nicely into the metric operators on the constituent ground
states.

\section{Haah's Cubic Codes}
\label{ap:haah} 
In this section, we provide a more detailed account of Haah's 17 cubic codes,
and the behavior of their ground states under pseudo-Hermitian perturbations. Throughout, we assume periodic boundary conditions as before. 

\begin{figure}
  \includegraphics[width=0.7\linewidth]{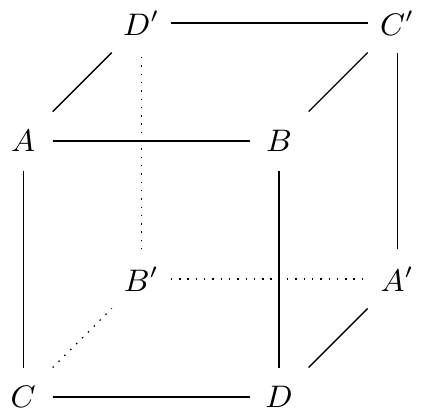}
  \caption{The stabilizers of Haah's cubic codes correspond to cube operators,
  with generically different operators on each vertex as labelled by $A$-$D$ and $A'$-$D'$. The operators at each
  vertex for the 17 different cubic codes are given in Table~\ref{table:cubicCodes}.}
  \label{fig:cubicStabilizer}
  \end{figure}

Haah's 17 CSS cubic codes~\cite{Haah2011} are defined on a cubic lattice,
with two Pauli spins on each vertex, $i$. There are two classes of stabilizers---one consisting solely of 
$Z$ operators, and the other with $X$ operators. The structure of these stabilizers is detailed in 
Fig.~\ref{fig:cubicStabilizer} and Table~\ref{table:cubicCodes}. In the polynomial representation used in Sec.~\ref{PolynomialRepresentation}, the 
stabilizers take the general form $Z(f,g)$ and $X(\bar{g}, \bar{f})$ for polynomials $f$ and $g$. As stated in the main text, these codes admit a large set of possible pseudo-Hermitian perturbations that leave
the code subspace real: in analogy to \equref{OneNaturalChoiceOfEta}, a very natural set of choices for the 
metric operator $\eta$ is given by \equref{EtasForTheHaahCodes}.

Since all the stabilizers in Haah's cubic codes are mutually commuting, all ground states have the same eigenvalue under $\eta$, provided $\eta$ can be assembled by stabilizers. For the toric code, the X-cube and checkerboard model discussed in the main text, it is straightforward both to find the combination of stabilizers that yield $\eta$ on a lattice with an even number of sites in all directions, and to show that $\eta$ cannot be made of stabilizers on any other lattice. For Haah's codes, the more complex form of the stabilizers makes the analysis more demanding, but possible using the polynomial representation of stabilizers \cite{macwilliams1977theory}. 

Using the same conventions as in Sec.~\ref{PolynomialRepresentation}, the metric operators in \equref{EtasForTheHaahCodes} can be written as 
\begin{align}\begin{split}
    \eta &= Z(h,h), \, X(h,h), \, iX(h,h)Z(h,h), \\
    h & = \sum_{j=1}^{L_x}\sum_{k=1}^{L_y}\sum_{\ell=1}^{L_z} x^{j-1}y^{k-1}z^{\ell-1}. \label{ExpressionOfEtaAppendix}
\end{split}\end{align}

We will first consider $\eta = Z(h,h)$. For stabilizers $Z(f,g)$, a choice of covering 
(i.e., a product of stabilizers at different points) can be specified by a \textit{covering 
polynomial} $k$, with the covering given by $Z(kf, kg)$. For example, if $k = 1 + x$, then the 
covering $Z(kf, kg)$ would consist of the product of two stabilizers---one at the origin,
and one at $(x,y,z) = (1,0,0)$. Therefore, the question of whether $\eta$ can be assembled
from stabilizers is equivalent to the question of whether $h = kf = kg$ for some polynomial
$k$. Mathematically, this factorization takes place in the quotient ring $P/I$, where $P$ is
the ring of polynomials of three variables with coefficients over $\mathbb{F}_2$, and $I$ is
the ideal generated by $x^{L_x} + 1$, $y^{L_y} + 1$, and $z^{L_z} + 1$. This quotienting
procedure imposes the periodic boundary conditions of the model.

We calculate this factorization with the computer algebra system SageMath. Generically,
this factorization procedure will yield two different coverings, $h = k_f f = k_g g$. 
To determine whether these two coverings are compatible, we calculate whether $k_f + k_g$
can be separated into two polynomials $d_f + d_g$, where $d_f \in (I:f)$ and
$d_g \in (I:g)$, where $(I:f)$ is the \textit{colon ideal}, $(I:f) = \{p \in P: pf \in I\}$. 
This is equivalent to checking whether $k_f + k_g$ belongs to the ideal generated by 
$(I:f) \cup (I:g)$. If such
a separation exists, then $k_f + d_f = k_g + d_g \equiv k$, and $h = kf = kg$ in $P/I$.
This covering may not be unique, as $k+d_{fg}$ also works as a covering, where $d_{fg} \in (I:f) \cap (I:g)$; 
however, for the purposes of understanding the behavior of non-Hermitian
perturbations, we are only interested in the existence of such a covering. We note that this procedure
should always be able to find a covering $k$ if it exists, so if a decomposition
$k_f + k_g = d_f + d_g$ does not exist, it should imply the non-existence of a covering.

Once we have obtained the covering $k$ for $Z(h,h)$, we immediately know that $X(h,h)$ in \equref{ExpressionOfEtaAppendix} can be assembled from $X$-stabilizers with the covering $\bar{k}$, since $X(\bar{k}\bar{g},\bar{k}\bar{f}) = X(\bar{h},\bar{h}) = X(h,h)$.

This calculation is done in SageMath 
for system sizes 
$L_x \times L_y \times L_z$ for $1 \leq L_x, L_y, L_z \leq 19$. Although the 
existence/non-existence of a covering follows no clear pattern for very small
system sizes, we see regular behavior emerge once the system size is larger than 
$3 \times 3 \times 3$. Specifically, the existence/non-existence of a covering
for a certain cubic code is only dependent on whether each length is even or odd
and, if it is even, whether it is divisible by $4$. This admits $3^3 = 27$ different
possible classes of system sizes---however, we find that some classes are equivalent
in terms of which cubic codes have coverings on them.
A full table of this behavior is shown in Table~\ref{table:cubicCoverings}. We note several
trends. On an odd-by-odd-by-odd lattice, none of the 17 cubic codes have code subspaces that
stay real under pseudo-Hermitian perturbations. If only a portion of the system lengths are odd, the reality of the code subspace 
depends on which dimensions have odd lengths, and whether the remaining lengths are divisible by 4.
In contrast, if $L_x$ and $L_y$ are divisible by $4$ and $L_z$ is even, all the code subspaces
stay real under pseudo-Hermitian perturbations.
Overall, cubic code 17 is the most unstable to pseudo-Hermitian perturbations, in that its code subspaces will become
complex for almost all system sizes. In contrast, cubic code 7 has the most stable code subspace. There are some groups of codes with the same
sensitivity to system sizes. If we consider codes with the same behavior up to a lattice rotation, these groups
are $(11, 12, 14, 15)$, $(5, 8, 10, 16)$, and $(2,3,6,9)$. It is interesting to note that, with the exception of cubic 
code 16, all codes within a group transform the same under entanglement renormalization~\cite{Dua2019}.
Note that cubic codes related by modular transformations~\cite{Dua2019a}---specifically, cubic codes 5/9 and 15/16---may have different system size dependencies. This is because a stabilizer covering in one model will generically transform non-trivially under modular transformations. In other words, the existence of a stabilizer covering in one cubic code does not imply the existence of a stabilizer covering in another cubic code related to the original by a modular transformation.

While our results are purely numerical, an analytic verification of these trends for all system sizes is
likely possible if one was to manually follow the factorization processes carried out in SageMath and
show that their conclusions are only sensitive to the system sizes' evenness/oddness and whether 
they are divisible by $4$. We do not attempt this, as there are $459$ separate cases that 
must be checked ($27$ possible system sizes for the $17$ codes), 
and instead analyze the numerical results which show clear trends
up to $19 \times 19 \times 19$ lattices.

  \begin{table}
  \begin{tabular}{c|c c c c c c c c}
       & $A$ & $B$ & $C$ & $D$ & $A'$ & $B'$ & $C'$ & $D'$ \\
       \hline
       \hline
      1 & $ZI$ & $ZZ$ & $IZ$ & $ZI$ & $IZ$ & $II$ & $ZI$ & $IZ$ \\
      2 & $IZ$ & $ZZ$ & $ZI$ & $ZI$ & $ZI$ & $ZZ$ & $IZ$ & $ZI$ \\
      3 & $IZ$ & $ZZ$ & $ZZ$ & $ZI$ & $ZZ$ & $II$ & $IZ$ & $IZ$ \\
      4 & $IZ$ & $ZZ$ & $ZI$ & $ZI$ & $IZ$ & $II$ & $IZ$ & $ZI$ \\
      5 & $ZI$ & $ZZ$ & $II$ & $ZZ$ & $ZI$ & $II$ & $IZ$ & $IZ$ \\
      6 & $ZI$ & $II$ & $ZI$ & $ZZ$ & $IZ$ & $ZZ$ & $II$ & $IZ$ \\
      7 & $ZI$ & $ZZ$ & $ZI$ & $IZ$ & $IZ$ & $II$ & $II$ & $ZZ$ \\
      8 & $ZI$ & $ZI$ & $IZ$ & $ZZ$ & $IZ$ & $II$ & $IZ$ & $ZI$ \\
      9 & $ZI$ & $IZ$ & $ZZ$ & $ZZ$ & $IZ$ & $ZZ$ & $II$ & $IZ$ \\
      10 & $ZI$ & $IZ$ & $ZI$ & $ZZ$ & $IZ$ & $ZZ$ & $ZI$ & $ZI$ \\ 
      11 & $ZI$ & $ZZ$ & $II$ & $IZ$ & $ZI$ & $II$ & $IZ$ & $ZZ$ \\
      12 & $ZI$ & $IZ$ & $ZZ$ & $ZZ$ & $ZI$ & $II$ & $II$ & $IZ$ \\
      13 & $ZI$ & $ZZ$ & $IZ$ & $ZI$ & $IZ$ & $II$ & $II$ & $ZZ$ \\
      14 & $ZI$ & $IZ$ & $ZZ$ & $ZZ$ & $IZ$ & $II$ & $ZZ$ & $IZ$ \\
      15 & $ZI$ & $IZ$ & $II$ & $ZZ$ & $IZ$ & $ZZ$ & $II$ & $ZI$ \\
      16 & $ZI$ & $ZI$ & $II$ & $IZ$ & $IZ$ & $ZZ$ & $II$ & $ZZ$ \\
      17 & $ZI$ & $ZZ$ & $IZ$ & $ZI$ & $IZ$ & $ZI$ & $ZI$ & $ZZ$
  \end{tabular}
  \caption{The $Z$ stabilizers for Haah's 17 CSS cubic codes, defined on the
  eight vertices of a cube, with vertices labeled according to Fig.~\ref{fig:cubicStabilizer}. The $X$ stabilizers are obtained by exchanging
  $A \leftrightarrow A'$, and likewise for the other vertices, and by exchanging
  the two Pauli spins on each site.}
      \label{table:cubicCodes}
  \end{table}

     \begin{table}
  \begin{tabular}{@{}  >{\centering\arraybackslash} m{1.67cm}|  *{8}{>{\centering\arraybackslash} m{0.7cm}}@{}}
       System Size & $CC_1$ & $CC_2$ \newline $CC_3$ \newline $CC_6$ \newline $CC_9$ & $CC_4$ & $CC_5$ \newline $CC_8$ \newline $CC_{10}$ \newline $CC_{16}$ & $CC_7$ & $CC_{11}$ \newline $CC_{12}$ \newline $CC_{14}$ \newline $CC_{15}$ & $CC_{13}$ & $CC_{17}$ \\
       \hline
       \hline
      E $\times$ E $\times$ E \\ E $\times$ E $\times$ e & \cmark & \cmark & \cmark & \cmark & \cmark & \cmark & \cmark & \cmark \\
      \hline
      e $\times$ E $\times$ E \\ o $\times$ E $\times$ E \\ E $\times$ e $\times$ E \\ E $\times$ e $\times$ e & \cmark & \cmark & \cmark & \cmark & \cmark & \cmark& \cmark & \xmark \\
      \hline
      e $\times$ e $\times$ E \\ e $\times$ E $\times$ E \\ E $\times$ o $\times$ E \\ e $\times$ e $\times$ e & \cmark & \cmark & \cmark & \cmark & \cmark & \cmark & \xmark & \xmark \\
      \hline
      E $\times$ E $\times$ o & \cmark & \cmark & \cmark & \cmark & \cmark & \xmark & \xmark & \cmark \\
      \hline
      e $\times$ o $\times$ E \\ E $\times$ o $\times$ e \\ e $\times$ o $\times$ e & \xmark & \cmark & \cmark & \cmark & \cmark & \cmark & \xmark & \xmark \\
      \hline
      e $\times$ E $\times$ o \\ E $\times$ e $\times$ o \\ e $\times$ e $\times$ o \\ e $ \times$ o $\times$ o \\ E $\times$ o $\times$ o & \xmark & \cmark & \xmark & \cmark & \cmark & \xmark & \xmark & \xmark \\
      \hline
      o $\times$ e $\times$ E \\ o $\times$ E $\times$ e \\ o $\times$ e $\times$ e & \xmark & \xmark & \cmark & \cmark & \cmark &\cmark  &\xmark & \xmark \\
      \hline
      o $\times$ o $\times$ E \\ o $\times$ o $\times$ e & \xmark & \xmark & \xmark & \xmark & \cmark & \cmark & \xmark &  \xmark \\
      \hline
      o $\times$ E $\times$ o \\ o $\times$ e $\times$ o \\ o $\times$ o $\times$ o & \xmark & \xmark & \xmark & \xmark & \xmark & \xmark & \xmark &  \xmark \\
  \end{tabular}
  \caption{The reality of the code subspace of Haah's cubic codes under pseudo-Hermitian perturbations are 
  highly sensitive to the system size. The reality of the subspace depends on whether each dimension length 
  is odd (o), even and divisible by 4 (E), or even and not divisible by 4 (e). Shown are all 17 of Haah's 
  cubic codes and the dependence of the code subspace stability on the system size. Codes with identical 
  dependencies have been grouped together, and some have been redefined by a spatial rotation. The trends listed 
  have been confirmed numerically to hold from system sizes $3 \times 3 \times 3$ to $19 \times 19 \times 19$.}
      \label{table:cubicCoverings}
  \end{table}
\clearpage
\bibliography{fracton}
\end{document}